\documentclass[%
floatfix,
reprint,
nofootinbib,
superscriptaddress,
amsmath,amssymb,
aps,
prb,
nomerge,
longbibliography,
showkeys,
frontmatterverbose
]{revtex4-2}
\usepackage{placeins}
\usepackage[T1]{fontenc}
\usepackage{newtxtext, newtxmath}
\usepackage{color}
\usepackage{graphicx}
\usepackage{dcolumn}
\usepackage{bm, bbm}
\usepackage{physics}
\usepackage{empheq}
\usepackage{hyperref}
\usepackage[capitalize]{cleveref}
\usepackage{enumitem}
\usepackage{xcolor}
\usepackage{orcidlink}
\usepackage[normalem]{ulem}
\usepackage{todonotes, soul}

\usepackage[byauthor, separator = {;\ }]{credits}

\begin{document}
	\preprint{APS/123-QED}
	\title{Discrete Time Crystal Order in Spin-Chains Enabled by Floquet Flat-Bands}	
	\author{Mahbub Rahaman\,\orcidlink{0000-0002-7455-2364}}
	\email[Primary \& Corresponding author:\;]{mahbub.phys@gmail.com}
	\credit{MR}{Conceptualization, Data curation, Project administration, Formal analysis, Investigation, Methodology, Resources, Software, Visualization, Writing -- original draft}
	\affiliation{Harish Chandra Research Institute, A CI of Homi Bhabha National Institute, Chhatnag Road, Jhunsi, Prayagraj, Uttar Pradesh 211019, India}
	\author{Analabha Roy\,\orcidlink{0000-0002-4797-0624}}
	\email[Co-Corresponding author:\;]{daneel@utexas.edu}
	\credit{AR}{Resources, Conceptualization, Writing -- review \& editing}
	\affiliation{Department of Physics, The University of Burdwan, Barddhaman, West Bengal 713104, India}
	
	\begin{abstract}
		We propose a novel protocol to realize discrete time-crystal (DTC) order in clean, periodically driven spin-$1/2$ chains. In each drive cycle, a global spin flip is followed by a two-tone flat-band segment. This flat-band segment engineers a fully degenerate Floquet quasienergy spectrum, suppresses thermalization, and stabilizes a robust period-doubled subharmonic response. Using exact time evolution, we identify a pronounced subharmonic peak at half the drive frequency in the Fourier spectrum of the order parameter, thereby providing clear evidence for the emergence of stable DTC. The resulting phase is insensitive to system size, interaction strength, and interaction range; however, it remains sensitive to spin-rotation errors ($\varepsilon_r$), which can destabilize the subharmonic response. Compared with disorder-induced many-body localized (MBL) and disorder-free dynamically many-body localized (DMBL) DTCs, we find that the exact flat-band protocol offers a broader tunability of drive parameters, whereas MBL and DMBL-based DTCs are more resistant to $\varepsilon_r$. In particular, the $\varepsilon_r$ sensitivity can be suppressed by incorporating additional spin-spin interactions that have modest deviations from the ideal flat-band protocol. This manifests itself in a robust DTC response over a finite window of spin-coupling strengths and drive frequencies. Our results establish flat-band driving as a versatile and experimentally relevant route to DTC order in disorder-free spin systems and motivate further exploration of nonequilibrium phases.
	\end{abstract}
	
	\maketitle
	
	\section{Introduction}
	Symmetry is a foundational principle in theoretical physics, underpinning quantum mechanics~\cite{Greiner_2012, Strocchi_2021, Bender_2024, Ziaeepour_2015, lombardi_2016, MACK1993361}, quantum field theory~\cite{Harlow2021, Bender_2021, Gripaios_2024}, the Standard Model~\cite{willenbrock_2005, Bernabeu2021, Robinson2014}, condensed-matter physics~\cite{Wu_2022, mcgreevy_2023}, and related areas~\cite{Yanofsky2017}. Many infinite dimensional quantum systems exhibit spontaneous symmetry breaking (SSB), in which ground states do not inherit the full symmetry of the Hamiltonian~\cite{Brauner_2024}. A wide range of conventional phases of matter can be understood through SSB, making it central to phenomena such as magnetism and superfluidity~\cite{LandauLifshitz_StatPhys, Goldenfeld_PT}, quantum Hall effects~\cite{Thouless_1982, goerbig2009}, and many-body localization that can protect or enable forms of symmetry breaking through athermality (eigenstate order)~\cite{Basko_2006, Nandkishore_2015, Khemani_2017_prx, Abanin_2019}.
	
	The SSB of translational symmetry is a core theme in equilibrium condensed-matter physics. For example, crystalline order arises from breaking the spatial translation symmetry. Analogously, in non-equilibrium dynamics, time-translation symmetry breaking (TTSB) refers to periodic dynamics that arises in the absence of explicit time-dependence~\cite{CTC1_2018, CTC2_2019} or in periodically driven systems where the quasistationary state does not share the periodicity of the drive~\cite{Sacha_2021}.
	Wilczek’s original proposal of such time crystals—spontaneous periodic motion in classical~\cite{Shapere_2012} and quantum~\cite{Wilczek_2012} equilibrium systems—was precluded by no-go theorems~\cite{Bruno_2013_1, Nozieres_2013, Watanabe_2015}. In the quest to obtain time crystals in closed quantum systems, this necessitated a change of focus to driven systems, where  time periodicity enables Floquet or discrete time crystals (DTCs)~\cite{Sacha_2015, Sacha_2021, Else_2016} that were to be subsequently realized in trapped ions~\cite{Zhang_2017}, superconducting qubits~\cite{Khemani_2021} and ultracold atoms~\cite{Choi2023}.
	
	In spin-1/2 chains, DTC order is associated with breaking the global $\mathbb{Z}_2$ symmetry and is diagnosed through an order parameter $\hat{\mathcal{O}}$ evaluated in the state of out-of-equilibrium $\ket{\psi(t)}$. Defining $\mathcal{S}(t)\equiv\expval{\hat{\mathcal{O}}}{\psi(t)}$, the hallmarks are: (a) TTSB, $\mathcal{S}(t)\neq \mathcal{S}(t+T)$ despite $\hat{H}(t)=\hat{H}(t+T)$; (b) rigidity, a robust subharmonic response with $\mathcal{S}(t)=\mathcal{S}(t+2T)$ without fine-tuning; and (c) persistence, stability of the dynamics of $2T$ over long times, including toward the thermodynamic limit.
	
	The DTC order is operationally characterized by the presence of a robust subharmonic peak in the Fourier spectrum of the order parameter $\hat{\mathcal{O}}$~\cite{Russomanno_2017}. However, the eigenstate thermalization hypothesis (ETH)~\cite{Deutsch_1991, Srednicki_1994} predicts that isolated quantum spin chains generically heat up and thermalize, leading to the eventual decay of the DTC order. Consequently, stabilization mechanisms are essential to maintain DTC phases. One prominent route is disorder-induced MBL~\cite{Basko_2006, Nandkishore_2015, Zhang_2017}, which suppresses heating by localizing many-body eigenstates and thus preserves memory of the initial state for long times. However, MBL stabilized DTC phases can remain vulnerable to internal excitations that trigger domain-wall proliferation and ultimately destroy the subharmonic response~\cite{Hauschild_2016}. Alternative stabilization routes include integrable spin systems~\cite{Roy_2025_1, Schaefer_2019}, clean driven systems~\cite{Rahaman_2024, Wing_2019, huang_2018, Anisur_2025, Sayan_2021, Lyu_2020}, and central-spin models~\cite{Hillol_2025, Hillol_2025_2}.
	
	Although earlier theoretical and experimental realizations of DTC order predominantly relied on disorder-induced localization as a heating-suppression mechanism~\cite{Else_2016, Khemani_2016, yao_2017, Randall_2021, Zhang_2017, Choi_2017}, an increasingly active direction has been the search for \emph{clean} (disorder-free) routes to stabilizing subharmonic responses. In periodically driven spin models, such stabilization can arise from dynamical mechanisms that strongly slow down or suppress thermalization even in the absence of quenched randomness~\cite{Pizzi_2021, Kozin_2019, Rahaman_2024}. A widely explored paradigm is drive-induced \emph{freezing}, in which the interplay of drive amplitude and frequency creates regimes where the stroboscopic evolution becomes nearly trivial for relevant initial states, thereby preventing rapid dephasing and thermalization~\cite{Das_exotic, Haldar_prx, Rahaman_2024_prb}. This phenomenon has been demonstrated across a range of clean systems, from effective two-level systems~\cite{Ashhab_2007} to interacting spin chains, including nearest-neighbor transverse-field Ising models (TFIM)~\cite{Das_exotic} as well as long-range spin-$1/2$ models~\cite{Rahaman_2024_prb}. In these cases, localization typically occurs only within structured regions of the drive-parameter space—often characterized by special values (``freezing points'') where the effective Floquet dynamics suppresses transitions and locks the system close to its initial configuration for long times. In long-range many-body systems (e.g., the Lipkin-Meshkov-Glick model), closely related effects can extend to disorder-free \emph{dynamical many-body localization} (DMBL)~\cite{Rahaman_2024_prb}, providing a qualitatively distinct mechanism for sustaining long-lived subharmonic order. These developments underscore the need for the suppression of thermalization in order to stabilize the DTC phases. 
	
	Furthermore, the transient nature of disorder-induced localization~\cite{Nandkishore_2015,mbldies_2022} motivates the exploration of more versatile and tunable mechanisms to realize discrete time crystals. The flat-band protocol, recently explored in this context~\cite{tista2025, Krishanu_2025}, offers a promising alternative by engineering a completely degenerate quasienergy spectrum through carefully articulated periodic drives. This engineered degeneracy localizes the system's initial state, thereby suppressing thermalization. Notably, this mechanism remains effective across a wide range of interaction strengths and ranges, as well as diverse drive parameters, providing a flexible platform for realizing the DTC order in various spin models. The novelty of our work lies in adapting this two-toned flat-band drive protocol to a periodically driven spin-1/2 chain, where we demonstrate the emergence of a stable and resilient DTC phase.
	
	We propose a protocol to realize a DTC phase in a clean, periodically driven spin-$1/2$ chain. Each driving cycle (with period $T=2\pi/\omega$ and frequency $\omega$) consists of a global spin-flip pulse followed by a flat-band segment. The latter is implemented via a two-tone drive that engineers a fully degenerate Floquet quasienergy spectrum, thus suppressing heating and stabilizing robust $\mathbb{Z}_2$ symmetry-breaking dynamics. Our exact time-evolution and Floquet numerical results demonstrate a stable $2T$--subharmonic response, characterized by a pronounced peak at $\omega/2$ in the Fourier spectrum of magnetization. However, on longer timescales we observe a departure from the ideal flat-band behavior, where spin interactions begin to significantly influence the system’s dynamics. This interaction-driven evolution steers the system toward a thermal regime, giving rise to a prethermal discrete time crystal (DTC).	We further quantify the robustness of the resulting phase against variations in spin-spin interaction strength and against spin-rotation errors. Compared with disorder-stabilized phases of MBL and DMBL induced DTCs, we found that flat-band-induced DTC is broadly independent of system size and interaction strengths or ranges, whereas MBL-based DTCs are comparatively more resilient to rotation errors. Finally, we show that incorporating an additional interaction term within the flat-band segment can substantially enhance the robustness of the DTC order in the presence of spin-rotation errors within specific regions of drive frequency and spin-interaction parameter space. These results identify the flat-band driving protocol as a practical route to DTC order in disorder-free spin systems and motivate the experimental exploration of such nonequilibrium phases.
	
	This article is organized as follows. Section~\ref{sec:model} introduces the model and the driven dynamics, emphasizing the flat-band construction and its role in stabilizing the DTC order. Section~\ref{sec:DTC} presents numerical evidence for the emerging DTC phase. Section~\ref{sec:dtc_stability} analyzes stability with respect to variations in spin interaction strengths, ranges, and spin rotational errors. Section~\ref{sec:dtc_stability_comparative_MBL_DMBL} compares the flat-band protocol with disorder-stabilized MBL and clean DMBL scenarios.  Section~\ref{sec:imperfect_flatband} explores the persistence of DTC against deviations from the ideal flat-band condition. Section~\ref{sec:conclusion} concludes with a summary and outlook.
	
	\section{The model and system dynamics}
	\label{sec:model}
	\begin{figure}[t!]
		\centering
		\includegraphics[width = 1.0\linewidth]{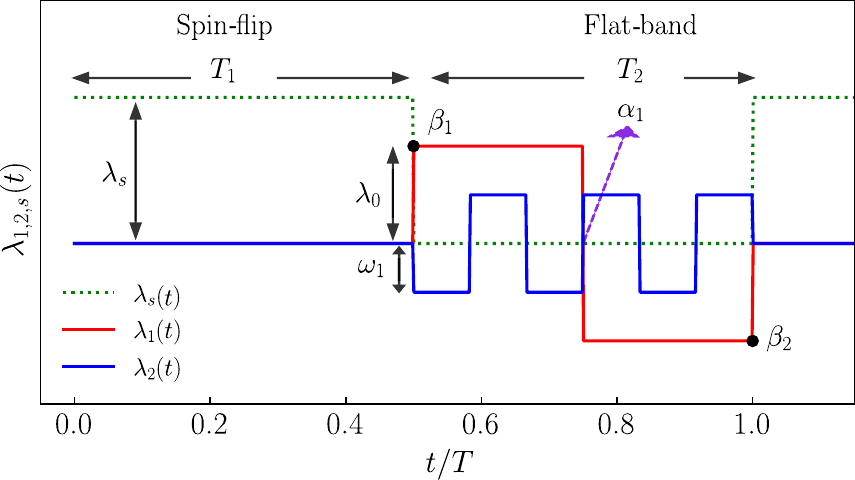}
		\caption{Three time dependent drives $\lambda_{1,2,s}(t)$ protocol for the proposed flat-band DTC model. The first time interval, $T_1$, consists of a spin-flip drive, while the second time interval, $T_2$, consists of a flat-band protocol.}
		\label{fig:flatband_drive_protocol}
	\end{figure}
	The generic flat-band protocol employs two-tone time dependent drives~\cite{Divakaran2014,Naji2022,Sau2014},  specifically designed to ensure that the drive signals are out of phase whenever they switch polarities. This causes unit propagation at integer multiples of the time period, thus inducing a flat-band in the Floquet energy spectrum and consequently nullifying the dynamics of the system~\cite{tista2025}. In principle, this protocol can be utilized to suppress heating and stabilize any subharmonic response to system and protocol defects. As a specific example, consider a one-dimensional spin-1/2 chain comprising $N$ sites, driven by two time-dependent Hamiltonians $\hat{H}_{1,2}(t)$, expressed as:		
	\begin{align}
		\label{eq:tc_protocol}
		\hat{H}(t) = 
		\begin{cases}
			\hat{H_1}(t), & 0\leq t < T_1,\\
			\hat{H_2}(t), & T_1\leq t < T.
		\end{cases}
	\end{align}
	Here, $T$ denotes the period of the complete drive protocol $\hat{H}(t)$, which is composed of two successive intervals: $\hat{H}_1(t)$ acts during the first interval $T_1$, whereas $\hat{H}_2(t)$ acts during the second interval $T_2$ ($T_2 = T - T_1$).  Unless otherwise stated,  $T_1 = T_2 = T/2$. In Eq.~\eqref{eq:tc_protocol}, the Hamiltonian $\hat{H_1}(t) $ is defined as 
	\begin{equation}
		\hat{H_1}(t) \equiv\; \hat{H}_{\mathrm{SF}}(t) = \lambda_s (1-\varepsilon_r) \hbar \sum^{N}_{i=1} \hat{\sigma}^x_i,
		\label{eq:spinflip:hamilt}
	\end{equation}
	where, $\hbar$ is the reduced Planck's constant. In addition, $\hat{\sigma}^{\mu=x,y,z}_i$ are the Pauli matrices at site $i$.
	Now, we consider the ideal case where $\varepsilon_r=0$~\footnote{Small values of $\varepsilon_r$ have been introduced during the stability analysis described in Sect.~\ref{sec:dtc_stability}}.
	If the value of $\lambda_s$ is set to $\omega/2$ (where $\omega=2\pi/T$), such that $\lambda_s T = \pi$, then the Hamiltonian $\hat{H}_1$ in Eq.~\eqref{eq:tc_protocol} simply flips each spin over the course of its duration. During this interval ($T_1$), the Hamiltonian $\hat{H}_1(t)$ does not incorporate any spin-spin interaction terms, thereby preserving the system's coherence. In the subsequent time interval, $T_2$, the system's dynamics is governed by the flat-band protocol, represented by the Hamiltonian $\hat{H_2}(t) \equiv \;\hat{H}_{\mathrm{FB}}(t) = \sum_{l=1,2} \lambda_l (t) \hat{\Lambda}_l$, as illustrated in Fig.~\ref{fig:flatband_drive_protocol}. The operators $\hat{\Lambda}_l$ are many-body in nature and satisfy the commutation relation $\comm{{\hat\Lambda_1}}{\hat{\Lambda}_2} \neq 0$. This protocol operates at dual rates, with the time-periodic functions $\lambda_l (t)$ having corresponding frequencies that are integer multiples of each other. The turning points at times $t_{1,2} = \beta_{1,2} T$, as depicted in Fig.~\ref{fig:flatband_drive_protocol} represent the instances where the Hamiltonian corresponding to flat band vanishes, i.e. $\hat{H}_{\mathrm{FB}}(\alpha_1 T) = 0$, with $\alpha_1 = (\beta_1 + \beta_2)/2$.
	
	A paradigmatic example of $\hat{H}_{\mathrm{FB}}$ is the transverse field Ising model (TFIM) with time-periodic fields, a prototypical example of an integrable spin chain~\footnote{We also consider nonintegrable long-range models in Sect.~\ref{sec:dtc_stability}}. The periodic fields are utilized to construct the two-rate induced flat-band. The Hamiltonian for the TFIM-driven flat band is expressed as:
	\begin{equation}
		\label{eq:fb_t}
		\hat{H}_{\mathrm{FB}}(t) = \lambda_1 (t) J \hbar\sum_{i=1}^{N} \hat{\sigma}_i^x \hat{\sigma}_{i+1}^x + \lambda_2 (t) h_z \hbar \sum^N_{i=1} \hat{\sigma}_i^z,
	\end{equation}
	with open boundary conditions (OBC).  Here, the time-dependent functions $\lambda_1(t)$ and $\lambda_2(t)$ are defined as
	\begin{align}
		\label{eq:two_drive_protocol_1}
		\lambda_1(t) &= +\lambda_0 && ;\frac{mT}{4} < t \le \frac{(m+1)T}{4},\nonumber\\
		&= -\lambda_0 && ;\frac{(m+1)T}{4} < t \le \frac{(m+2)T}{4},\nonumber\\
		\lambda_2(t) &= \omega_0 - \omega_1 && ;\frac{qT}{4p} < t \le \frac{(q+1)T}{4p},\nonumber\\
		&= \omega_0 + \omega_1 && ;\frac{(q+1)T}{4p} < t \le \frac{(q+2)T}{4p},
	\end{align}
	where $m, q$ are even natural numbers ($m, q = 0, 2, 4 \dots$), and $p \in \mathbb{Z}$. Here, $J$ represents the strength of the nearest-neighbor spin coupling, and $h_z$ denotes the amplitude of the transverse field. Two time-dependent functions ($\lambda_{1,2}(t)$) with distinct amplitudes and frequencies are engineered to effectively suppress the system dynamics at the end of the drive-cycle, thereby producing a pure degenerate flat-band (see Appendix~\ref{app:fb}). This protocol ensures that the flat-band protocol is independent of the drive parameters, such as the drive frequency ($\omega$) and the spin coupling strength ($J$), a unique feature that is contingent upon both the flat-band protocol itself and the boundary conditions of the spin chain. In our numerical simulations we use open boundary conditions (OBC) for the spin interactions, although periodic boundary conditions (PBC) may also be used and are often more natural for spin chains. The qualitative dynamics under PBC and OBC are similar, and the robustness of the DTC phase does not depend on the choice of boundary conditions. A  detailed comparison between OBC and PBC on the spin-1/2 chain to obtain DTC under flat band is provided in Appendix~\ref{app:boundary_condition}. 
	
	For our numerical and analytical investigations, we set the primary drive frequency $\omega$ of the time-dependent Hamiltonian $\hat{H}_{1,2}(t)$ to $\omega = 20$. The additional default parameters are $p=3$, $\hbar = 1$, $\lambda_s = \omega/2 = 10$, $h_z = 1$, $\omega_0 = 0$, $\omega_1 = \lambda_0 J/2$, $J = 1$, and $\lambda_0 = 0.18$, as illustrated in Fig.~\ref{fig:flatband_drive_protocol}. These parameters ensure consistency with previous work~\cite{yao_2017} and yield a minimal width of the Floquet spectrum~\cite{tista2025}.
	\begin{figure}[t!]
		\centering
		\includegraphics[width = 1.0\linewidth]{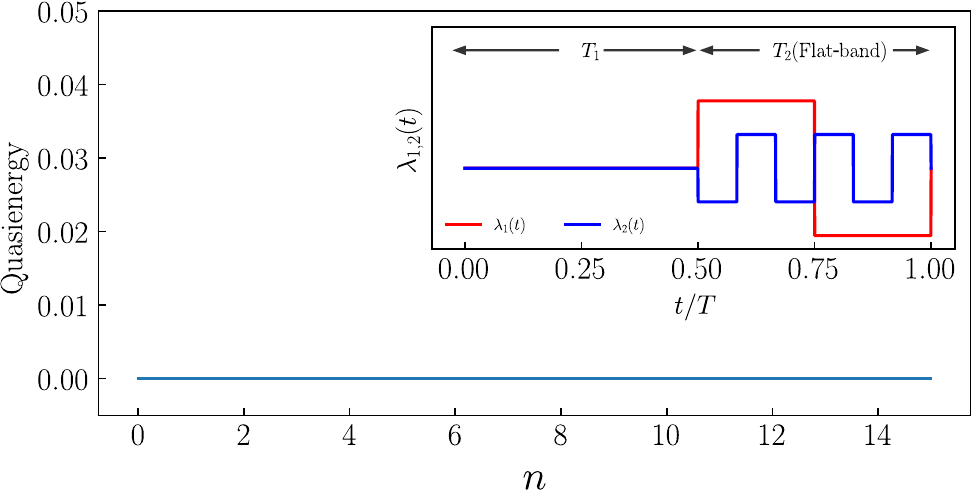}
		\caption{quasienergy spectrum of the proposed flat-band protocol within the DTC model. quasienergy levels plotted against the quasienergy eigenstates (n). The quasienergy levels exhibit all-degeneracy at zero energy, indicating the emergence of the flat band. The corresponding $\hat{H}_{\mathrm{FB}}(T_2)$ drive protocol is illustrated in the inset.}
		\label{fig:flatband_spectrum}
	\end{figure}
	
	To validate the emergence of this flat-band numerically, we have obtained and analyzed the quasienergy spectrum. Specifically, we set $N=8$ spins and simulated the time-evolution described in Eq.~\eqref{eq:tc_protocol}. For this analysis, we disable the spin-flip drive, \textit{i.e.}, set $\hat{H}_1 = 0$ during the $T_1$ interval. This ensures that the dynamics is solely governed by the flat-band protocol $\hat{H}_{\mathrm{FB}}(t)$ that is activated during the $T_2$ interval.  The quasienergy spectrum is obtained by diagonalizing the Floquet operator ($\hat{\mathcal{F}}$) at time $T$ using the QuTiP module (Quantum Toolbox in Python~\cite{qutip}). According to Quantum Floquet theory~\cite{shirley_1965,sambe_1973,Holthaus_2016,mbeng_2024}, the time-evolution operator for a time-periodic Hamiltonian over one period $T$, given by $\hat{\mathcal{F}} = \mathcal{T} \exp\left[-i \int_0^T \hat{H}(t) dt/\hbar\right]$, where $\mathcal{T}$ denotes time-ordering, produces eigenvalues that can be expressed as $e^{-i \epsilon_\alpha T/\hbar}$, where $\epsilon_\alpha$ are the quasi-energies defined modulo $\omega$. The corresponding eigenstates, known as Floquet modes, satisfy $\hat{\mathcal{F}} \ket{\phi_\alpha} = e^{-i \epsilon_\alpha T/\hbar} \ket{\phi_\alpha}$. The quasienergy spectrum provides information on the long-time dynamics of periodically driven systems.	
	
	The quasienergy spectrum exhibits complete degeneracy, with all levels collapsing to zero energy, as illustrated in Fig.~\ref{fig:flatband_spectrum}. This confirms that the flat-band protocol effectively localizes the spin-chain wave function and can suppress thermalization. Upon activating $\hat{H}_1$ and repeatedly applying the complete drive protocol described in Eq.~\eqref{eq:tc_protocol}, the spins are expected to return to their initial configuration at even multiples of the time period, thereby exhibiting period-doubling behavior. This periodic recurrence violates the intrinsic discrete time-translation symmetry of the driven system, giving rise to a DTC phase with $2T$ periodicity. The following section presents a detailed numerical investigation of this emergent DTC phase.
	
	\section{Emergence of the DTC}
	\label{sec:DTC}
\begin{figure}[t!]
		\centering
		\includegraphics[width = 1.0\linewidth]{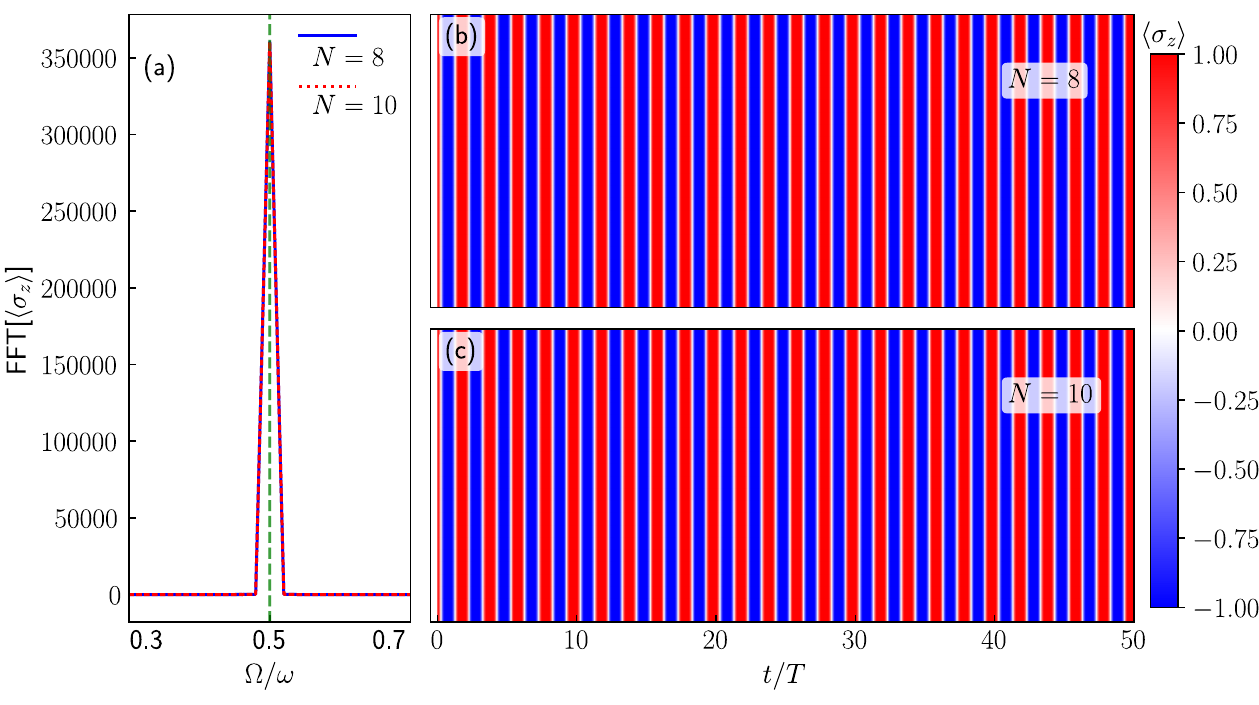}
		\caption{Temporal evolution of the magnetization $\expval{\hat{\sigma}^z(t)}$ for the proposed DTC model in spin chains of sizes $N=8$ and $10$, with the spin rotational error fixed at $\varepsilon_r = 0$. The magnetization oscillates within the range [-1, 1] at each discrete time period, demonstrating a clear 2T periodicity, as depicted in (b) and (c). (a) illustrates the FFT of the magnetization, highlighting a distinct subharmonic peak at $\omega/2$ for both system sizes, thereby confirming the realization of the DTC phase.}
		\label{fig:dtc_magnetization}
	\end{figure}

	We numerically simulated the complete drive protocol described in Eq.~\eqref{eq:tc_protocol} for long times, with $\hat{H}_{\mathrm{FB}}$ set to Eq.~\eqref{eq:fb_t}. Specifically, we ran two instances comprising $N=8$ and $N=10$ spins, respectively. In both instances, the spin chain was initially prepared in a fully spin-polarized state. Additionally, we turn off the spin rotation error ($\varepsilon_r = 0$), while keeping all other parameters consistent with Sect.~\ref{sec:model}. Simulations were performed to evaluate the propagator in QuTiP, and the results were utilized to compute the time evolution of the order parameter which, in this case, is the magnetization $\displaystyle \expval{\hat{\sigma}^z(t)} = \expval{\hat{\sigma}^z}{\psi(t)}$. As shown in Fig.~\ref{fig:dtc_magnetization}(b-c), the magnetization displays a clear $2T-$periodicity, although the underlying primary drive is $T-$periodic. This corroborates the breaking of discrete time-translation symmetry and confirms the emergence of a DTC phase in the spin chain.
	To further corroborate the emergence of the DTC phase, we compute the Fourier spectrum of the magnetization $\expval{\hat{\sigma}^z(t)}$ over an extended time window, as illustrated in Fig.~\ref{fig:dtc_magnetization}(a). The spectrum exhibits a pronounced subharmonic peak at $\Omega = \omega/2$, without any other discernible features, thus unambiguously confirming the $2T$-periodic nature of the dynamics. The absence of additional spectral peaks, together with the nonextensive behavior of both the magnetization and its Fourier transform, can be attributed to the complete quasienergy degeneracy engineered by the flat-band protocol. This engineered degeneracy not only ensures the temporal resilience of the DTC phase, but also underpins its persistence over a larger system size ($N = 10$), underscoring the scalability of the proposed mechanism.
	
	\begin{figure}[t!]
		\centering
		\includegraphics[width = 1.0\linewidth]{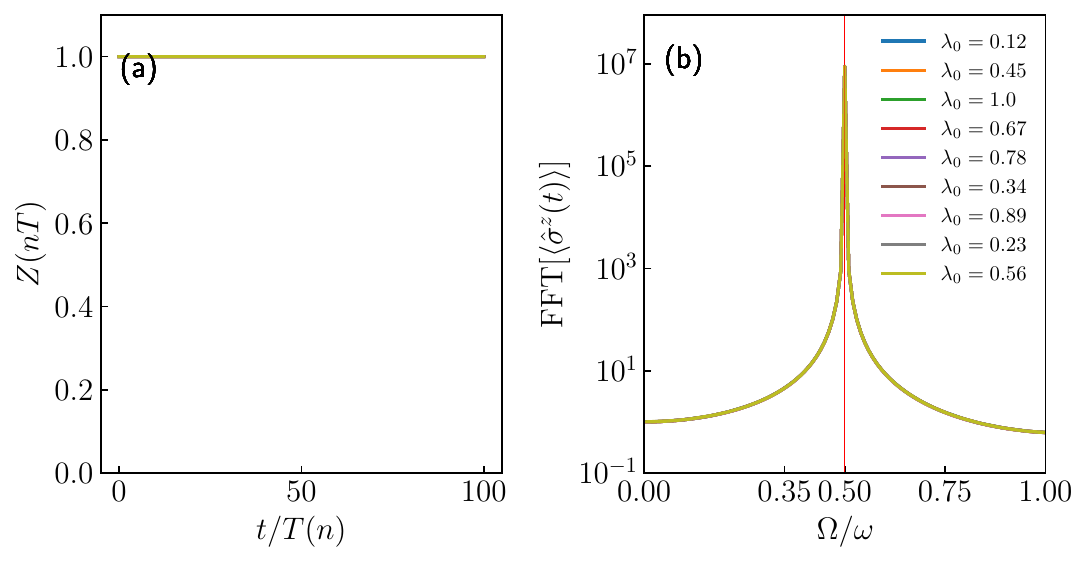}
		\caption{(a) illustrates the temporal variation of the parity-adjusted magnetization order parameter, $Z(nT)\equiv (-1)^n\expval{\hat{\sigma}^z(nT)}$ for the proposed DTC model in a spin chain of size $N=8$, subjected to varying spin-spin interaction strengths. (b) presents the FFT of the corresponding magnetization($\langle\hat{\sigma}^z (t) \rangle$), revealing the presence of a subharmonic peak at $\omega/2$ for all values of $\lambda_0$.}
		\label{fig:dtc_magnetization_lambda0}
	\end{figure}
	\section{Stability and Rigidity of the FB-DTC}
	\label{sec:dtc_stability}
	In the preceding sections, we demonstrated the emergence of the DTC phase stabilized with Floquet flat-bands, provided that the spins and the underlying system are prepared under ideal conditions. These conditions include the absence of spin rotational errors (i.e., $\varepsilon_r = 0$), a DTC drive that is  tuned to perform the spin-flip operations, a perfectly two-toned exact flat-band protocol, and the exclusive presence of nearest-neighbor spin-spin interactions with  coupling strength ($\lambda_0 J = 0.18$) as specified in Eq.~\eqref{eq:fb_t}. To evaluate the robustness of the proposed model under more realistic scenarios, we now simulate deviations from these ideal conditions and investigate the stability of the DTC phase.
	
	We begin by investigating the stability of the DTC phase with respect to variations in the \emph{spin-spin interaction strength} that range from very small to relatively large values. Numerical simulations are performed for different values of $\lambda_0 J\in [0.01, 1]$, while keeping all other parameters consistent with those outlined in Sec.~\ref{sec:model}, specifically $\omega = 20$, $\varepsilon_r = 0$, $h_z = 1$, $\omega_0 = 0$, $\omega_1 = \lambda_0 J/2$, and $\lambda_s = \omega/2 =10$. The temporal evolution of parity-adjusted magnetization order parameter, defined as $\displaystyle Z(nT) \equiv (-1)^n \expval{\hat{\sigma}^z}{\psi(nT)}=(-1)^n \langle \hat{\sigma}^z (nT) \rangle$ is computed and presented in Fig.~\ref{fig:dtc_magnetization_lambda0}. The results reveal that the 2T-DTC phase emerges robustly across the entire range of spin-spin interaction strengths and remains stable over extended timescales. This highlights the resilience of the DTC phase to variations in spin-spin interaction strength and underscores the effectiveness of the flat-band protocol in suppressing ETH~\cite{Deutsch_1991,Srednicki_1994,Rahaman_2024_prb,tista2025}.
	
	\begin{figure}[t!]
		\centering
		\includegraphics[width = 1.0\linewidth]{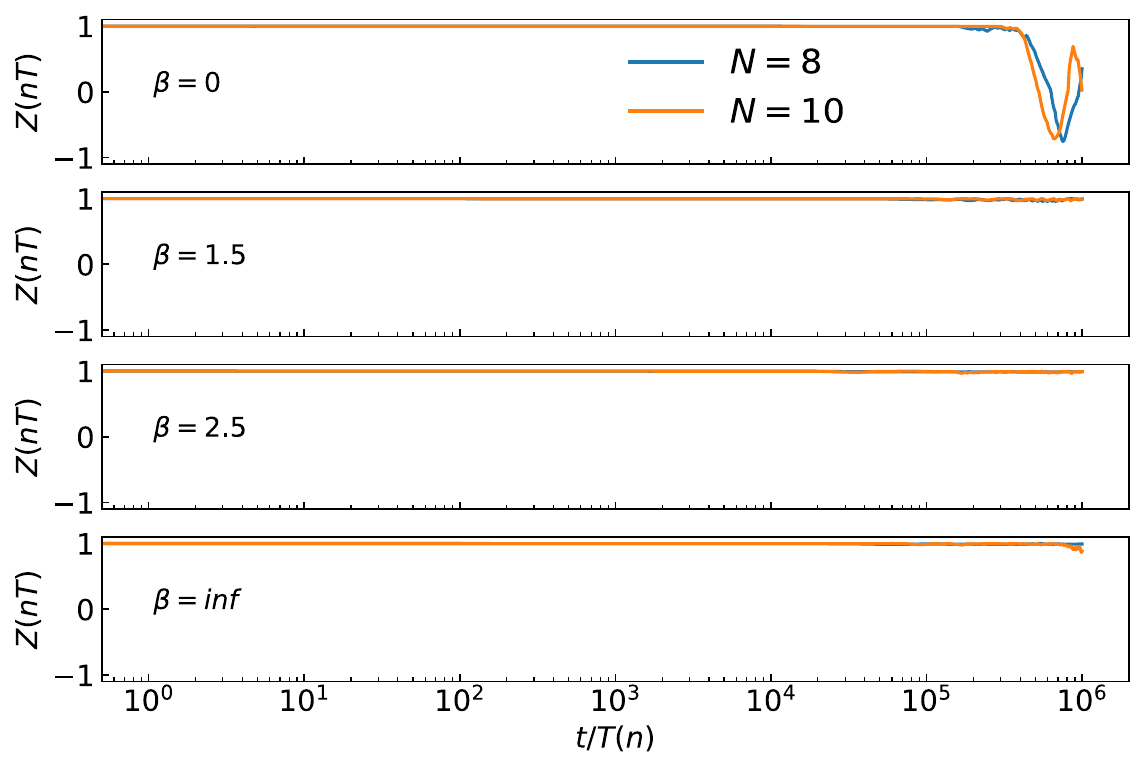}
		\caption{Temporal evolution of the magnetization order parameter, $Z(nT) = (-1)^n \expval{ \hat{\sigma}^z(nT)} $, at stroboscopic times for system sizes $N = 8, 10$ and various spin-spin interaction ranges ($\beta = 0, 1.5, 2.5, \infty(\text{inf})$).}
		\label{fig:dtc_magnetization_duty_cycle}
	\end{figure}
	Furthermore, we have investigated the stability of the DTC phase across various \emph{spin-spin interaction ranges}, characterized by the power-law decay form $J_{ij} = J/|i-j|^\beta$ for the $(i,j)^{th}$ spin-pair. Here, the flat-band Hamiltonian is modified from the paradigmatic example in Eq.~\eqref{eq:fb_t} to 
	\begin{equation}
		\label{eq:fb_t:longrange}
		\hat{H}_{\mathrm{FB}}(t) \to \lambda_1 (t)  \hbar\sum_{i< j} J_{ij}\hat{\sigma}_i^x \hat{\sigma}_j^x + \lambda_2 (t) h_z \hbar \sum_i \hat{\sigma}_i^z,
	\end{equation}
	Here, $\beta$ denotes the interaction-range parameter. Following Buyskikh’s benchmarking~\cite{buyskikh_2016}, we consider physically distinct regimes of $\beta$: $\beta = 0$ (all-to-all interactions), $\beta = 1.5$ and $\beta = 2.5$ (long-range interactions), and $\beta = \infty$ (inf), corresponding to nearest-neighbor interactions, i.e., the paradigmatic case in Eq.~\eqref{eq:fb_t}. We performed numerical simulations for spin-chain sizes $N = 8, 10$ and computed the parity-adjusted order parameter $\displaystyle Z(nT) \equiv (-1)^n \expval{\hat{\sigma}^z}{\psi(nT)}$ over long timescales (up to $10^6\,T$) in these interaction regimes. The results, shown in Fig.~\ref{fig:dtc_magnetization_duty_cycle}, indicate that the DTC phase emerges across all interaction ranges and remains robust for long durations in both system sizes. However, at later times, the order parameter exhibits a gradual decay in the all-to-all case ($\beta=0$), with a slightly faster decay rate than in the shorter-range cases ($\beta=1.5, 2.5, \infty$). This trend suggests comparatively stronger thermalization effects in the all-to-all regime, although overall decay remains weak and the DTC response persists on extended timescales for all $\beta$. At asymptotically long times, the system exhibits a gradual drift toward thermal equilibrium. This behavior reflects the deviation from the flat band, arising from spin–spin interactions, and is consistent with the presence of a prethermal discrete time crystal (DTC) phase within the proposed model. 
	
\begin{figure}[t!]
		\centering
		\includegraphics[width = 1.0\linewidth]{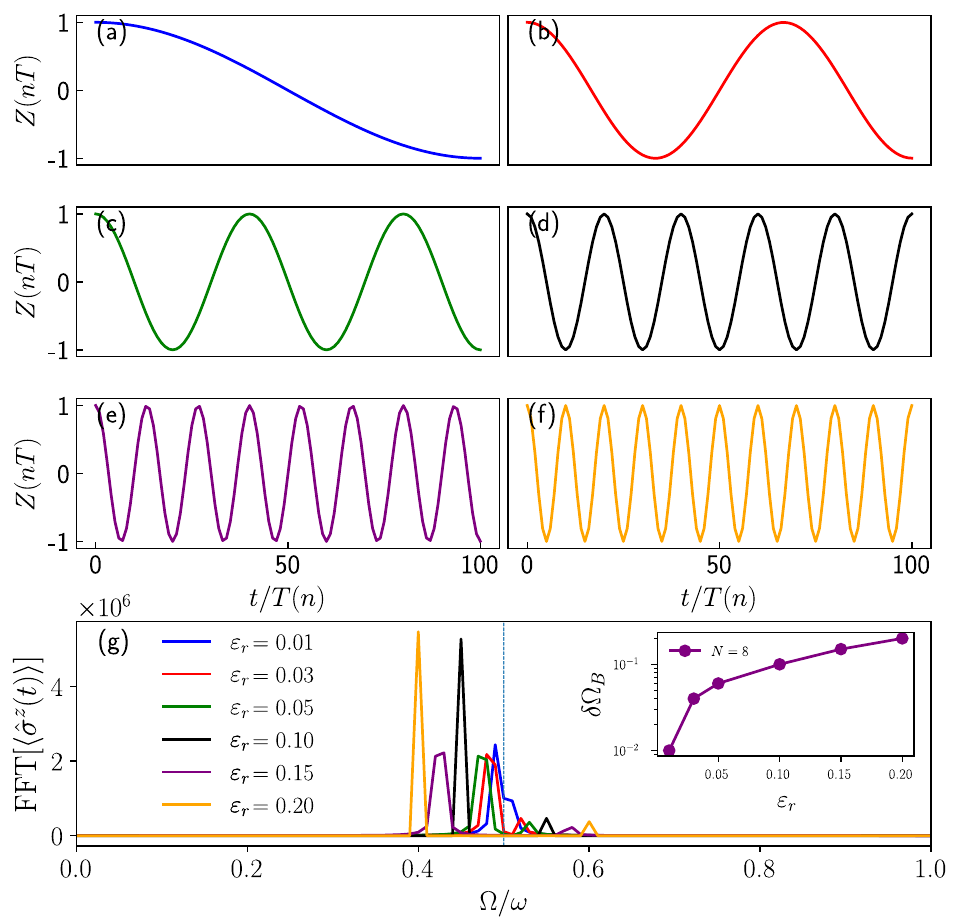}
		\caption{(a–f) display the temporal evolution of the magnetization order parameter, $Z(nT) \equiv (-1)^n\expval{\hat{\sigma}^z(nT)}$ for a spin chain of $N=8$ sites under systematically increasing spin-rotation errors ($\varepsilon_r = 0.01, 0.03, 0.05, 0.1, 0.15, 0.2)$. With growing $\varepsilon_r$, the initially perfect period-doubled oscillations gradually develop beating patterns, indicating the appearance of spectral components beyond the primary subharmonic frequency $\Omega = \omega/2$. (g) shows the corresponding FFT spectra, where well-resolved sidebands emerge in the vicinity of $\Omega = \omega/2$ (marked by a dotted blue vertical line). As illustrated in the inset, the beat frequency $\delta\Omega_B$ increases with $\varepsilon_r$, offering a quantitative signature of the progressive melting of the DTC phase induced by finite spin-rotational imperfections.}
		\label{fig:dtc_magnetization_error}
	\end{figure}

	To further assess the stability of the DTC phase, we introduce a small finite \emph{spin rotational error} ($\varepsilon_r > 0$) into the Hamiltonian $\hat{H}_1$, as defined in Eq.~\eqref{eq:spinflip:hamilt}. Numerical simulations are performed for $\varepsilon_r \in \{0.01, 0.03, 0.05, 0.1, 0.15, 0.2\}$ with the spin chain size fixed at $N=8$, while all remaining parameters are maintained at the values specified in Sect.~\ref{sec:model}. The temporal evolution of magnetization, presented in Fig.~\ref{fig:dtc_magnetization_error}(a--f), reveals the emergence of characteristic beat patterns for each nonzero rotational error, signifying the presence of spectral components beyond the subharmonic frequency $\Omega = \omega/2$. To identify these additional frequencies, we performed an FFT of the magnetization time data; the resulting spectra are shown in Fig.~\ref{fig:dtc_magnetization_error}(g). The FFT reveals distinct sidebands in the vicinity of $\Omega = \omega/2$, which are responsible for the observed beating in the magnetization dynamics. As shown in the inset of Fig.~\ref{fig:dtc_magnetization_error}(g), the beat frequency $\delta\Omega_B$ increases with $\varepsilon_r$, reflecting a progressive destabilization of the DTC phase. Collectively, these results underscore the sensitivity of the flat-band-induced DTC phase to spin-rotational imperfections, highlighting the importance of precise pulse-angle control for maintaining a stable subharmonic response.
	
	\section{Comparative Robustness of DTC Phases: Flat band Versus MBL/DMBL}
	\label{sec:dtc_stability_comparative_MBL_DMBL}	
	We have investigated the robustness of the flat-band (FB)-induced DTC order in relation to two widely studied localization mechanisms: disorder-induced MBL and drive-induced DMBL in a clean system. Using numerical simulations, we assess the stability of DTC through time-resolved magnetization and Fourier analysis, with particular focus on the beat frequency $\delta\Omega_B$ corresponding to the subharmonic peak. The following subsections present the comparative results and highlight the parameter regimes in which each mechanism remains the most resilient to spin-rotation errors and variations in interaction range.
	
	\subsection{Stability comparison of FB-DTC with disorder MBL-DTC}
	\label{sec:dtc_stability_flatband_mbl}
	We have investigated the robustness of the DTC phase induced by the flat-band protocol in comparison to that of the DTC phase stabilized via MBL. To that end, we follow the seminal work of Yao et al.~\cite{yao_2017}, which provides a comprehensive investigation into the emergence and stability of the MBL stabilized DTC phase in a one dimensional spin-1/2 chain. The Hamiltonian governing the MBL induced DTC is given by Eq.~\eqref{eq:tc_protocol} with the flat-band Hamiltonian modified from the paradigmatic example in Eq.~\eqref{eq:fb_t} to the following.
	\begin{equation}
		\hat{H}_{\mathrm{FB}} \to \hat{H}_{\mathrm{MBL}} = \sum_{i=1}^{N} J^{\prime}\hat{\sigma}^z_i  \hat{\sigma}^z_{i+1} + \sum_{i=1}^N B^z_i \hat{\sigma}^z_i.
		\label{eq:tc_protocol_13}
	\end{equation}
	Here, the random field $B^z_i$ is sampled from a uniform distribution within the range $[0, W]$, where $W = 2\pi$ represents the strength of the disorder, consistent with previous studies~\cite{yao_2017}. Additionally, the spin coupling strength is fixed at $J^{\prime} = 0.18$ to ensure the localization of the system and the stability of the DTC phase, as indicated by its level spacing ratio ($\expval{r} \approx 0.38$)~\cite{yao_2017, Khemani_2016}. Next, we investigate the dynamics governed by the drive protocols outlined in Eqs.~(\ref{eq:tc_protocol} and \ref{eq:tc_protocol_13}), with disorder strength $B^z_i$ uniformly distributed in the range $[0, 2\pi]$. Simulations are conducted for spin chain sizes $N = 8, 10, 12, 14$, with the DTC drive parameters set to default values. To assess the impact of spin rotational errors, we introduce a set of errors, $\varepsilon_r$, into the Hamiltonian $\hat{H}_{\mathrm{SF}}(t)$, as defined in Eq.~\eqref{eq:spinflip:hamilt}. The temporal evolution of magnetization $\expval{\hat{\sigma}^z(t)}$ is numerically computed from $t=0$ to $300T$, and averaged over $960$ disorder realizations. Additionally, we performed an FFT of the magnetization data to identify the dominant frequencies and quantify the associated beat frequencies ($\delta \Omega_B$). For comparison, we also present results from an analysis of the aforementioned quantities for the DTC induced by the flat-band protocol as described in Eq.~\eqref{eq:tc_protocol} for the same set of rotational errors. 
	\begin{figure}[t!]
		\centering
		\includegraphics[width = 1.0\linewidth]{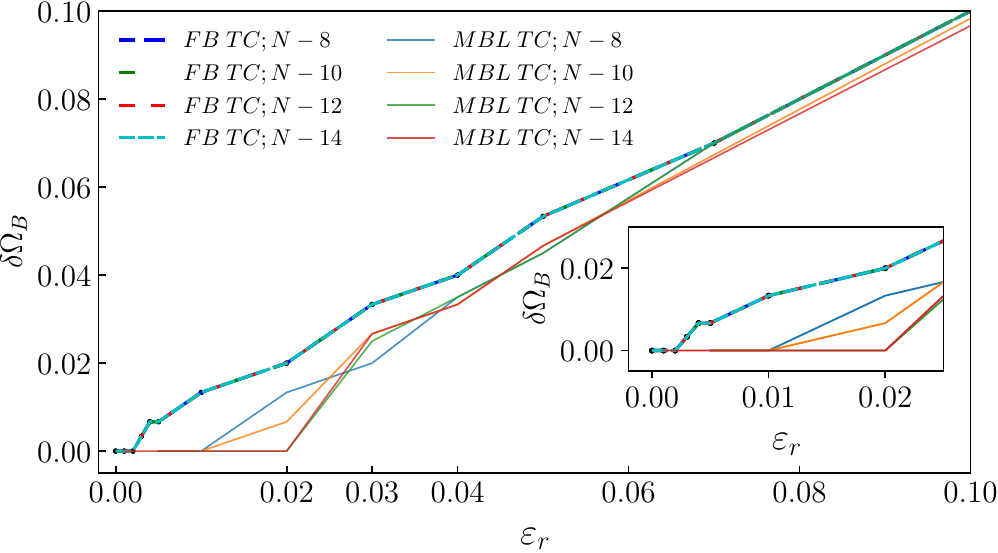}
		\caption{Comparison of the beat frequency of the magnetization $\expval{\hat{\sigma}^z(t)}$ for the FB-DTC (dashed curves) and the MBL-DTC phase (solid curves, averaged over 960 disorder realizations) across different spin chain sizes $N=8, 10, 12$ and $14$. The magnetization is evaluated over a duration of $300T$. The beat frequency ($\delta \Omega_B$) is plotted as a function of the spin rotational error ($\varepsilon_r$). The inset provides a magnified view of the main plot for enhanced clarity.}
		\label{fig:beat_magnetization_mbl_FB}
	\end{figure}
	
	The results of this comparative analysis are presented in Fig.~\ref{fig:beat_magnetization_mbl_FB}, revealing significant differences in the variation of $\delta \Omega_B$ between the FB-DTC and the MBL-DTC. The FB-DTC phase appears to be resilient to spin rotation errors up to $\varepsilon_r \approx 0.002$, beyond which $\delta \Omega_B$ increases linearly with $\varepsilon_r$. In contrast, the MBL-DTC phase demonstrates resilience up to $\varepsilon_r \approx 0.01$, followed by a regime (up to $\varepsilon_r \sim 0.02$) where $\delta \Omega_B$ decreases with increasing spin chain length. For larger $\varepsilon_r$, the beat frequency begins to increase linearly, approaching the scaling observed for the FB-DTC phase. These findings indicate that the FB stabilized DTC phase is more susceptible to spin rotational errors than the MBL stabilized DTC phase. The inset of Fig.~\ref{fig:beat_magnetization_mbl_FB} provides a magnified view of the main plot for improved clarity. Thus, the MBL-DTC phase exhibits superior robustness against spin rotational errors relative to the FB-DTC phase.
	
	\subsection{Stability Comparison of FB-DTC with DMBL-DTC}
	\label{sec:dtc_stability_flatband_localization}
	We now extend our investigation beyond the basic comparison with MBL-DTC to comparative studies of the stabilization of the flat-band DTC phase with DTC-phases obtained in clean systems where emergent localization mechanisms arise intrinsically from the dynamics via dynamical many-Body localization. In the context of spin chains, dynamical localization can give rise to localized states that resist thermalization, thereby providing a robust mechanism for stabilizing DTC phases. 
	
	To that end, we consider a spin-1/2 chain governed by the DMBL induced DTC protocol detailed in Eq.~\eqref{eq:tc_protocol} with
	the flat-band Hamiltonian modified from the paradigmatic example in Eq.~\eqref{eq:fb_t} to the following.
	\begin{equation}
		\hat{H}_{\mathrm{FB}} \to \hat{H}_{\text{DMBL}}(t) = \sum_{i<j} J_{ij}\hat{\sigma}^y_i  \hat{\sigma}^y_{j} + \lambda_{22}(t)  \sum_{i=1}^N \hat{\sigma}^z_i,
		\label{eq:tc_protocol_17}
	\end{equation}
	\begin{figure}[t!]
		\centering
		\includegraphics[width = 0.9\linewidth]{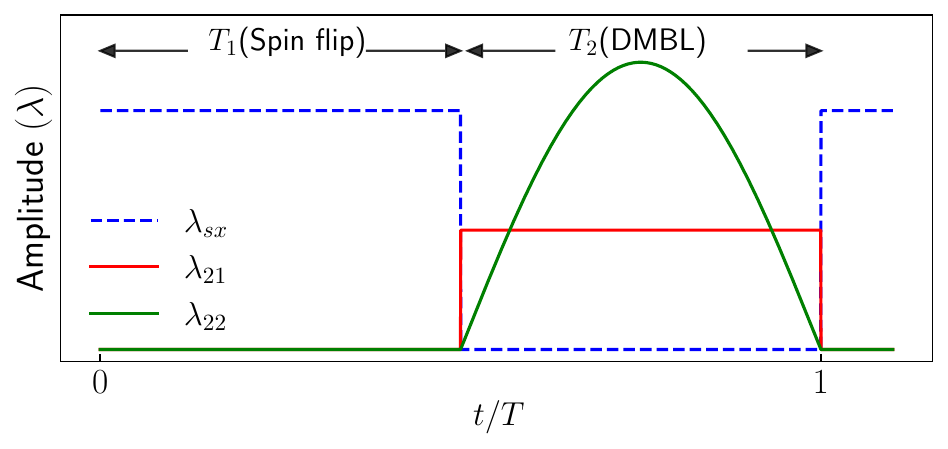}
		\caption{Schematic representation of the drive protocol for the proposed dynamical localization-induced DTC model in a clean spin-1/2 system. The protocol consists of two time-dependent drives: a spin-flip drive ($\lambda_{sx}$) applied during the first time interval $T_1$, followed by a time-dependent transverse field Ising model (TFIM) or Lipkin-Meshkov-Glick (LMG) model ($\lambda_{21}$) driven by a sinusoidal field with amplitude $\lambda_{22}$ during the second time interval $T_2$.}
		\label{fig:drive_DMBL}
	\end{figure}
	with power-law decay over the range $\beta$ in the spatial interaction amplitudes $J_{ij} = {J_0}/{\abs{i-j}^\beta}$, and spin coupling $J_0$. The physically distinct ranges of $\beta$-are classified according to Buyskikh's benchmarking detailed in Sect.~\ref{sec:dtc_stability}. The transverse field drive $\displaystyle \lambda_{22}(t) \equiv h \sin(\omega t)$ is characterized by a time-dependent amplitude modulated by the drive frequency $\omega$ and the amplitude $h$, as illustrated in Fig.~\ref{fig:drive_DMBL}. Consequently, the time-dependent Hamiltonian $\hat{H}_2(t)$ implemented during the second time interval $T_2$ 
	comprises the TFIM for $\beta = \infty$ and the LMG model for $\beta = 0$~\cite{Rahaman_2024_prb}. Note that the drive protocol during the $T_1$-cycle remains unchanged from the default protocol described in Eq.~\ref{eq:tc_protocol}, thereby preserving the recurrence of spin-flips and maintaining TTSB. Crucially, when the drive parameters $h$ and $\omega$ are tuned such that the ratio $\kappa \equiv 4\omega/h$, coincides with a root of the zeroth-order Bessel function of the first kind, such that $\mathcal{J}_0(\kappa) = 0$, the system becomes dynamically localized in its initial state throughout the second time interval $T_2$~\cite{Das_exotic,Rahaman_2024_prb,Rahaman_2024}. This condition, commonly referred to as the \emph{freezing point}, is characteristic of the high-frequency drive regime. In the low-frequency regime, on the contrary, no such localization occurs, and the system fails to retain memory of its initial state.

\begin{figure}[t!]
		\centering
		\includegraphics[width = 1.0\linewidth]{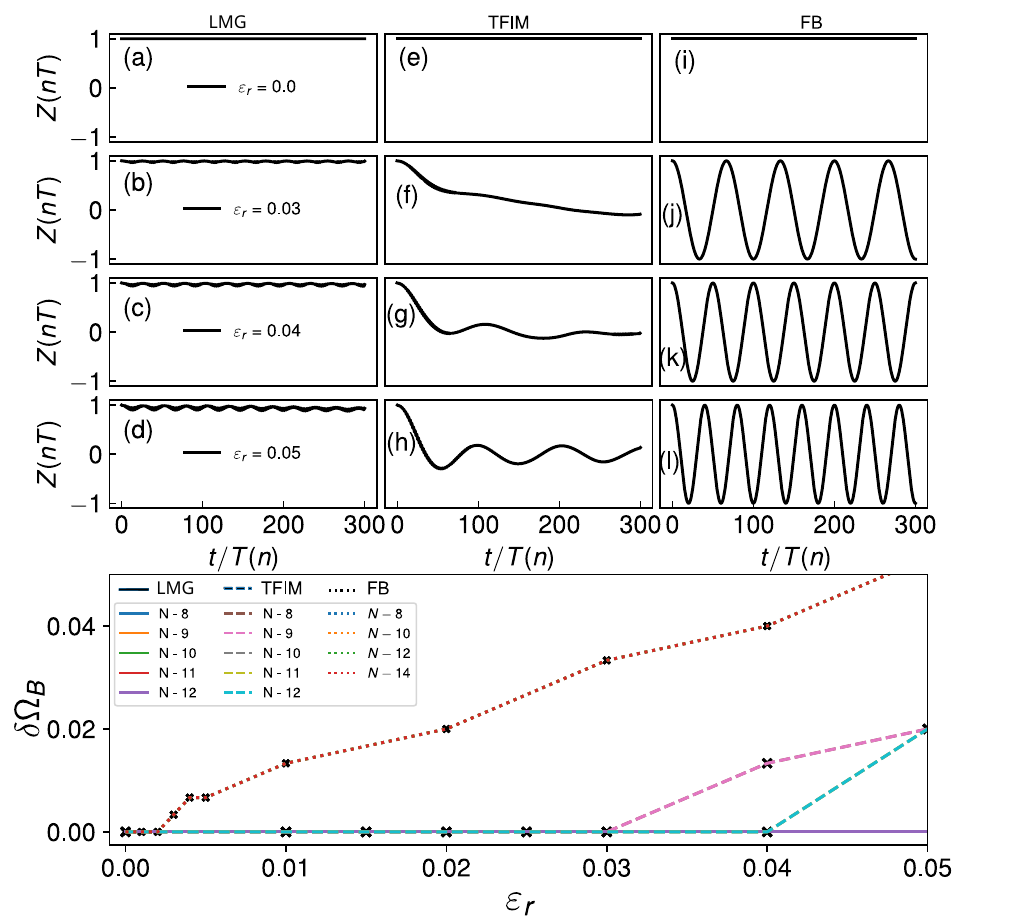}
		\caption{Temporal evolution of magnetization for different spin rotational errors $\varepsilon_r = (0.0, 0.03, 0.04, 0.05)$ for DMBL : LMG [(a)-(d)] and TFIM [(e)-(h)] and [(i)-(l)] flat-band localized DTCs over long timescales up to $300T$ for a spin chain of size $N=9$. The DMBL-induced DTC phase remains robust against $\varepsilon_r$, whereas the flat-band induced DTC phase exhibits faster melting as $\varepsilon_r$ increases. The bottom panel shows the beat frequencies ($\delta \Omega_B$) around $\Omega=\omega/2$ for different $\varepsilon_r$ values, calculated from the magnetization data for various spin chain lengths. While no $\delta \Omega_B$ is observed for the DMBL-induced DTC phase (solid or dashed lines), finite $\delta \Omega_B$ emerges for the FB-induced DTC phase ( dotted lines) when $\varepsilon_r \ge 0.002$.}
		\label{fig:magnetization_DMBL_TFIM}
	\end{figure}

	We investigate the stability of the DTC phase facilitated by dynamical localization in a spin chain of size $N=9$, considering both the TFIM ($\beta = \infty$) and the LMG model ($\beta = 0$). The spin chain is initialized in a fully polarized up-spin configuration. The drive parameters are set to $\omega = 20$ and $J_0 = 0.18$, with the drive amplitude ($h$) tuned to correspond to the freezing points of the system. Additionally, spin rotational errors ($\varepsilon_r = 0.0, 0.01, 0.02, 0.03, 0.04, 0.05$) are introduced during spin-flip operation in the first time interval $T_1$. The spin-flip drive is applied for a duration of $T_1 = T/2$, followed by the TFIM or LMG Hamiltonian, localized at freezing point during the remaining interval $T_2 = T/2$. Numerical simulations are conducted to compute the temporal evolution of parity-adjusted magnetization $\displaystyle Z(nT) \equiv (-1)^n \expval{\hat{\sigma}^z(nT)} = (-1)^n\expval{\hat{\sigma}^z}{\psi(nT)}$. Figure ~\ref{fig:magnetization_DMBL_TFIM} shows that magnetization exhibits $2T$ periodicity despite the primary drive being $T$ periodic, thus confirming the emergence of a DTC phase in the spin chains. The LMG-DTC phase is extremely robust against finite spin rotational errors, whereas the subharmonic exhibits  minor sensitivity for the TFIM-DTC case in small (finite) system sizes. This difference highlights the role of nonintegrability in localization: the integrable TFIM model is more susceptible to perturbations, while the nonintegrable LMG model maintains localization more robustly. The beat frequencies around $\Omega=\omega/2$ are obtained from the FFT decompositions for sizes ($N = 8, 9, 10, 11$ and $12$) (see Fig.~\ref{fig:magnetization_DMBL_TFIM}). The LMG-induced DTC phase does not exhibit any beats around $\Omega=\omega/2$ even for comparatively larger rotation errors (up to $\varepsilon_r=0.05$). In contrast, the TFIM-DTC phase exhibits beats for finite $\varepsilon_r \geq 0.03$ in systems with $N = 9$. However, for larger system sizes, the beat frequency $\delta \Omega_B$ decreases at finite $\varepsilon_r$, indicating enhanced DTC stability with increase in system size. In other hand the beat frequency of the flat-band-induced DTC phase begins to rise even for relatively small $\varepsilon_r \gtrsim 0.002$ for all system sizes (see bottom panel in Fig~\ref{fig:magnetization_DMBL_TFIM}), while the LMG- and TFIM-DTC phases remain stable under similar conditions. This indicates that the flat-band-induced DTC phase is more sensitive to finite spin rotational errors compared to the DMBL-induced DTC phases.
	
	\section{Persistent DTC under Imperfect flat bands}
	\label{sec:imperfect_flatband}
	
	The flat-band protocol introduced in Sect.~\ref{sec:model} is designed to achieve perfect localization: the two-tone drive is carefully tuned so that the effective Floquet operator reduces to the identity at integer multiples of the time period, thereby freezing all dynamics (see Appendix~\ref{app:fb}). In practice, however, precise realization of such a protocol is challenging, and small deviations from the ideal drive parameters are inevitable. Rather than merely degrading performance, such imperfections can, in certain cases, even confer additional stability to the system. To explore both effects, we examine two types of departure from the ideal flat band: (i) small perturbations in the frequencies of the two time-dependent flat-band drives, and (ii) the inclusion of an additional spin-spin interaction term in the faster-rotating component of the two-tone drive.
	
	\subsection{Deviation in the frequencies in the perfect flat band} 
	\noindent To model the above imperfections, we consider modified flat-band drive parameters expressed as:
	
	\begin{align}
		\label{eq:two_drive_protocol_1new}
		\lambda_1(t) &= +\lambda_0 && ;\frac{mT^{\prime}}{4} < t \le \frac{(m+1)T^{\prime}}{4},\nonumber\\
		&= -\lambda_0 && ;\frac{(m+1)T^{\prime}}{4} < t \le \frac{(m+2)T^{\prime}}{4},\nonumber\\
		\lambda_2(t) &= - \omega_1 && ;\frac{qT^{\prime \prime}}{4p} < t \le \frac{(q+1)T^{\prime \prime}}{4p},\nonumber\\
		&= + \omega_1 && ;\frac{(q+1)T^{\prime \prime}}{4p} < t \le \frac{(q+2)T^{\prime \prime}}{4p},
	\end{align}
	Here, $\delta \Omega_1 = $ and $\delta \Omega_2$ represent small deviations in the drive frequencies such that $T^{\prime} = 2\pi/(\omega + \delta \Omega_1)$ and $T^{\prime \prime} = 2\pi/(p\omega + \delta \Omega_2)$. All other parameters are kept consistent with the default values specified in Sect.~\ref{sec:model}. To quantify the impact of these imperfections, we calculate the fidelity of the system's state, defined as $F(2nT)$, where:
	\begin{equation}
		F(t) = \abs{\ip{\psi(0)}{\psi(t)}}^2
	\end{equation}
	Here, $\ket{\psi(0)}$ denotes the initial state, and $\ket{\psi(t)}$ represents the state of the system at stroboscopic times corresponding to integer multiples of the double period, $t = 2nT$ (with $n = 1, 2, 3, \dots$)~\cite{Rahaman_2024}. A high fidelity value ($F(t) \equiv 1$) indicates that the system remains close to its initial state, serving as a key metric to evaluate the robustness of the DTC phase.
	\begin{figure}[t!]
		\centering
		\includegraphics[width = 1.0\linewidth]{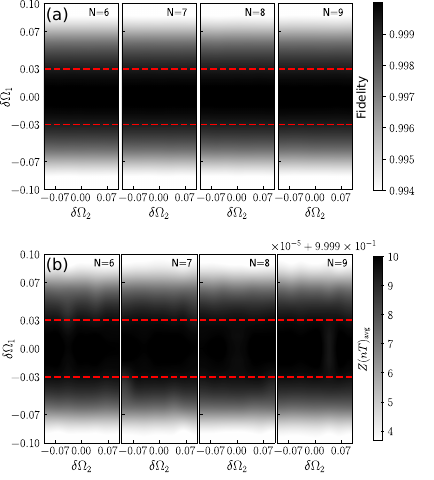}
		\caption{(a) Fidelity of the DTC phase as a function of deviations $\delta \Omega_1$ and $\delta \Omega_2$ for system sizes $N=6, 7, 8$ and $9$. The fidelity remains high within the range of small deviations (bounded by the red lines), demonstrating the robustness of the DTC phase against imperfections in the drive frequencies. (b) Time-averaged parity adjusted magnetization $Z(nT)_{\text{avg}}$ across the range of deviations $\delta \Omega_1$, $\delta \Omega_2$ and the system sizes. The results show that the order parameter remains robust against small deviations, consistent with the fidelity analysis.}
		\label{fig:fidelity}
	\end{figure}
	
	To systematically assess the impact of these imperfections, we numerically compute the fidelity for spin chains of sizes $N = 6, 7, 8, 9$ over a duration of $100T$, sweeping the deviations $\delta \Omega_1$ and $\delta \Omega_2$ across a range of values. As shown in Fig.~\ref{fig:fidelity}(a), the fidelity remains close to unity when the deviations are small ($\delta \Omega_1, \delta \Omega_2 \in [-0.03, 0.03]$), confirming that the DTC phase tolerates minor imperfections in the drive frequencies without appreciable degradation. However, as the deviations grow in magnitude, the fidelity gradually decreases, signaling a progressive loss of stability in the DTC phase. Additionally, we numerically calculate the time integrated parity adjusted magnetization, $ Z(nT)_{\text{avg}} = \frac{1}{n_{\text{max}}}\sum_{n=0}^{(n_{\text{max}}-1)} (-1)^n\langle \hat{\sigma}^z (nT)\rangle$ for $100T$ with same set of deviations $\delta \Omega_1$, $\delta \Omega_2$ and system sizes. The results, shown in Fig.~\ref{fig:fidelity}(b), reveal that the time-averaged order parameter remains robust against small deviations, consistent with the fidelity analysis. These observations paint a nuanced picture: while the flat-band protocol possesses an intrinsic resilience to practical imperfections, maintaining optimal performance still demands careful control over the drive parameters. In other words, the protocol is forgiving up to a point, but not unconditionally.
	
	\subsection{Introduction of additional spin-interaction term}
	\label{sec:add_interaction}
	To further investigate the robustness of the flat-band-induced DTC in the presence of spin-rotational errors under imperfect flat-band conditions, we introduce an \emph{additional aperiodic spin-spin interaction} along the order parameter direction (z-direction) into the two-tone flat-band Hamiltonian and examine its influence on the emergent localization properties of the spin chain. Specifically, we consider a modified time-periodic flat-band Hamiltonian of the form:
	\begin{align}
		\tilde{H}_{\mathrm{FB}}(t) &= \lambda_1(t) \sum_i \hat{\sigma}^x_i \hat{\sigma}^x_{i+1} + \lambda_2(t) \sum_i \hat{\sigma}^z_i + \lambda_f J\sum_i \hat{\sigma}^z_i \hat{\sigma}^z_{i+1},
		\label{eq:fb_interaction}
	\end{align}
	where the time-dependent coefficients are defined the same as in Eq.~\eqref{eq:two_drive_protocol_1}.
	Additionally, $\lambda_f$ represents the additional term for the spin-spin interaction strength and time coefficients are fixed at $m, q = 0, 2,4\dots$ and $p = 3$ for numerical simulations.
	To analyze the dynamics, we decompose the time evolution operator over one period into five intervals and apply the Magnus expansion:
	\begin{equation}
		\hat{U}(T, 0) = \exp(-i \hat{H}_{\mathrm{ME}} T),
	\end{equation}
	where $\hat{H}_{\mathrm{ME}}$ is the effective Hamiltonian expanded as a series in powers of the drive frequency. The effective Hamiltonian (see Appendix~\ref{app:interaction_flatband} for detailed derivations) is given by:
	\begin{equation}
		\hat{H}_{\mathrm{ME}} \equiv \frac{\lambda_f J}{2} \sum_i \hat{\sigma}^z_i \hat{\sigma}^z_{i+1}+ \frac{\lambda_s(1-\varepsilon_r)}{2} \sum_i \hat{\sigma}^x_i + \mathcal{\hat{O}}(T).
	\end{equation}
	The first term represents the additional spin-spin interaction, the second term corresponds to the spin-flip operation, and the third term accounts for the leading-order correction arising from their interplay.
	
	To analyze the DTC dynamics under this modified flat-band protocol, we investigate the exact DTC-Hamiltonian in different regimes of drive frequency and interaction strength. For strong interactions ($\lambda_0 J \sim 1$) and high-frequency driving ($\omega \sim 20$), the correction term becomes negligible, allowing the effective Hamiltonian to be dominated by the additional spin interaction and spin-flip terms. This ensures robust localization and a stable subharmonic response, even in the presence of spin-rotation errors. In contrast, in the low-frequency regime ($\omega \sim 1$), the last term becomes significant, introducing complex dynamics that can destabilize the DTC phase. However, for weaker interactions ($\lambda_0 J \sim 0.1$), the spin-spin interaction term becomes negligible under high-frequency driving, making the system more susceptible to spin-rotation errors. However, in the low frequency regime, the last term remains small due to the smaller amplitudes despite the low frequency. The additional spin interaction term effectively becomes large due to the small frequency and counteracts the additional spin rotational error, eventually obtaining robust DTC.

	To develop physical intuition for the interplay between rotational errors and the additional spin-spin interaction in stabilizing the DTC phase, we analyze the system dynamics using the toggling frame formalism (derivations are provided in Appendix~\ref{app:interaction_flatband}). In this frame, the periodic global spin-flip operation is absorbed into a time-dependent basis transformation, whereby local operator signs are periodically inverted and the $z$-direction coupling is strategically redistributed across the drive cycle~\cite{Choi_2017_prl}. Within this framework, the additional $\sigma^z_i\sigma^z_{i+1}$ interaction term contributes to the effective Floquet Hamiltonian through both averaged (zeroth-order) and higher-order commutator corrections, arising from the noncommutativity of sequential pulse intervals.
	
	In the high-frequency, strong-interaction regime (\(\lambda_0 J\sim 1\), \(\omega\gtrsim 20\)), the toggling-frame expansion exhibits parametric suppression of higher-order terms, rendering the leading-order effective Hamiltonian as a competition between the spin-flip term and the time-averaged interaction. Critically, the spin-rotation error \(\varepsilon_r\) appears only as a first-order perturbation to these dominant contributions, and its accumulated effect over a single period remains negligible. Consequently, the subharmonic (period-doubled) response persists with minimal degradation, and the DTC phase remains robust.
	
	Conversely, in the low-frequency regime (\(\omega\sim 1\)), the higher-order toggling-frame corrections become comparable to leading-order terms and are no longer suppressed. These higher-order corrections can induce complex mixing between the ideal flat-band evolution and undesired error-induced rotations, which generically weakens or destabilizes the subharmonic response. However, a compensatory mechanism emerges for weaker bare interactions (\(\lambda_0 J\ll 1\)) at low frequencies: the averaged $z$ coupling becomes relatively more significant and its contribution can systematically counteract the accumulated phase errors from spin rotations, thereby suppressing drift in the effective dynamics. In this regime, the additional spin-spin interaction acts as a stabilizing agent against rotational perturbations. Overall, the DTC stability depends sensitively on the competition between interaction strength and drive frequency; depending on their relative magnitudes, the additional spin-spin term can either enhance or undermine DTC persistence in the presence of finite rotational errors. This mechanism provides a coherent explanation for the diverse robustness behaviors observed in the numerical simulations presented next.
	
	\begin{figure}[t]
		\centering
		\includegraphics[width = \linewidth]{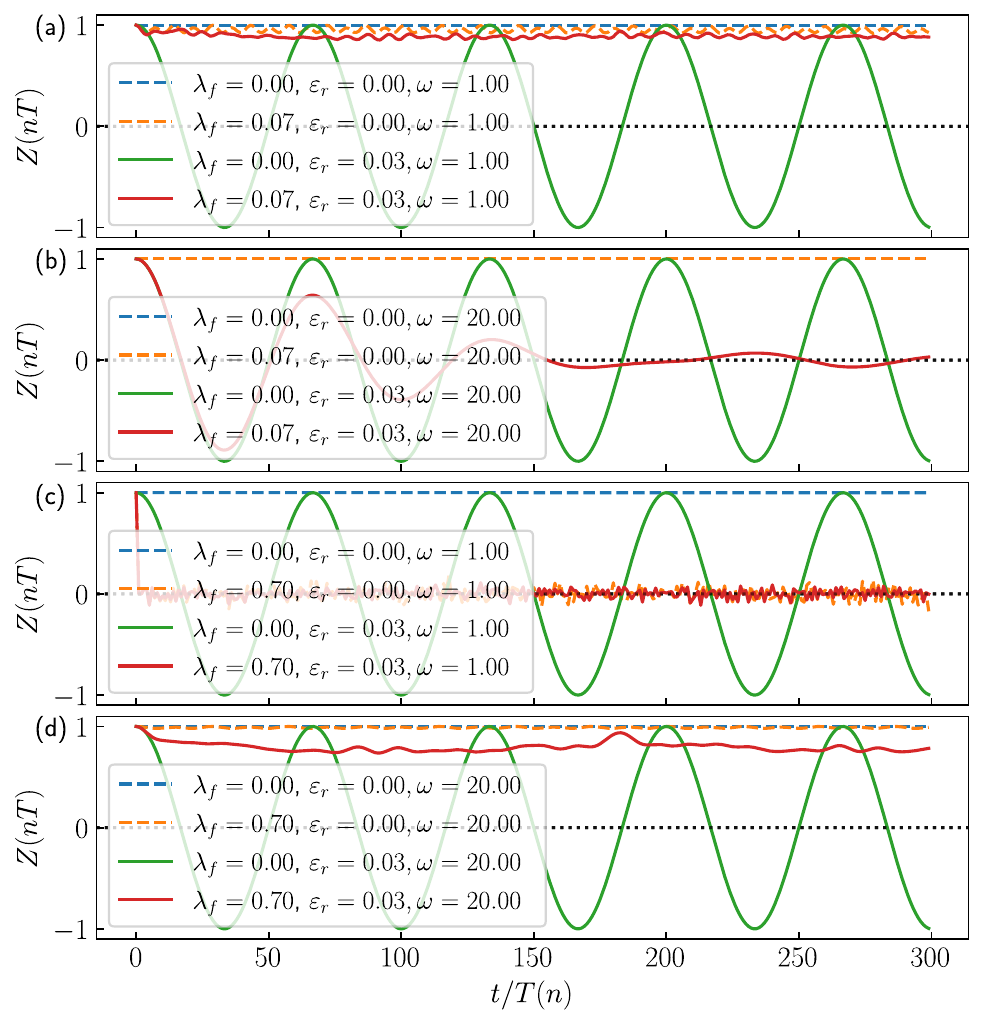}
		\caption{Temporal evolution of the magnetization order parameter $Z(nT)$ under the modified flat-band protocol incorporating an additional spin-spin interaction term, evaluated for various values of $\lambda_{0, f}$, $\omega_0$, $\varepsilon_r$, and drive frequency $\omega$. The results demonstrate that the DTC phase exhibits resilience to spin-rotation errors($\varepsilon_r = 0.03$) in regimes characterized by strong interactions and high-frequency ($\omega=20$) driving (d), as well as in regimes with weak interactions and low-frequency($\omega=1$) driving (a).}
		\label{fig:perfect_flatband}
	\end{figure}
	We investigate this argument numerically by simulating the time evolution of the parity adjusted magnetization order parameter $Z(nT) \equiv (-1)^n \langle \hat{\sigma}^z_i(nT) \rangle$ under the modified flat-band protocol for various values of 	$\lambda_f J = 0.7\lambda_0J$, $J=1$, $\omega_0 = 0$ for $\lambda_0 \in \{1, 0.1\}$, and $\omega \in \{1, 20\}$, in the presence of spin-rotation error ($\varepsilon_r = 0.03$) 
	for a spin chain of size $N=8$. The results, depicted in Fig.$~\ref{fig:perfect_flatband}$, reveal that the DTC phase remains robust against spin-rotation errors in regimes characterized by strong interactions and high-frequency driving, as well as in regimes with weak interactions and low-frequency driving. Therefore, the flat-band protocol with an additional spin-spin interaction term can stabilize the DTC phase under specific regimes of drive frequency and interaction strength.

    Thus, the stabilization follows from the structure of the effective Floquet Hamiltonian. In the ideal flat-band protocol, the leading-order effective Hamiltonian vanishes, so the system has no intrinsic dynamical energy scale to suppress imperfections. Adding interactions generates a finite effective Hamiltonian in the toggling frame, creating quasienergy splittings that prevent this coherent buildup, producing the observed enhancement of DTC stability. Importantly, interactions are not the primary mechanism behind the DTC order. The flat-band protocol alone yields stable subharmonic oscillations without imperfections. Interactions instead enhance robustness to pulse errors by generating a finite effective dynamical energy scale that suppresses their coherent accumulation. This role differs fundamentally from interactions in DMBL time crystals: there, interactions are essential to stabilize the phase, whereas here they are secondary, serving only to mitigate imperfections.

	\begin{figure}[t]
		\centering
		\includegraphics[width = 1.0\linewidth]{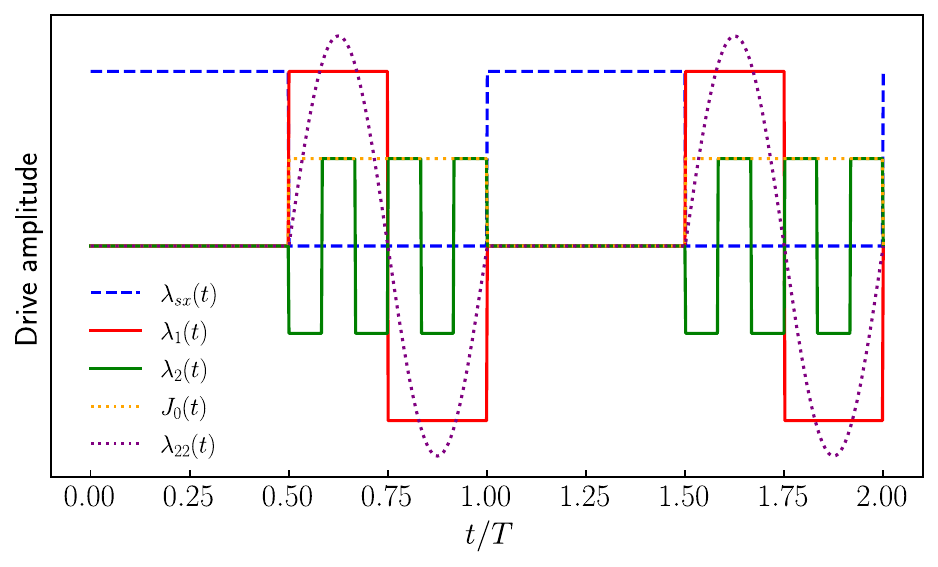}
		\caption{Schematic representation of the mixed protocol combining the flat-band drive and the DMBL drive. The protocol consists of two time intervals: $T_1=T/2$ for the spin-flip operation and $T_2= T/2$ for the mixed flat-band ($\lambda_1 (t)$ and $\lambda_2 (t)$) and DMBL ($J_0(t)$, $\lambda_{22}(t)$) drives.}
		\label{fig:mixed_protocol}
	\end{figure}
	\subsection{DTC emerging from mixed flat-band protocol and DMBL}
	\label{app:mixed_FB_DMBL}
	
	We have observed an enhancement of DTC stability in the presence of an additional interaction term $\lambda_f J$ under the flat-band protocol. To further investigate the stability of the DTC phase in the presence of rotational errors, we develop a \emph{mixed protocol} that combines the flat-band protocol with a dynamical many-body localization (DMBL). The mixed protocol is defined as follows:

	\begin{align}
		\label{eq:tc_protocol1}
		\hat{H}_{\text{mixed}}(t) = 
		\begin{cases}
			\hat{H_1}(t), & 0\leq t < T_1,\\
			\hat{H_2}(t), & T_1\leq t < T,
		\end{cases}
	\end{align}
	which is similar to the two-tone drive protocol defined in Eq.~\eqref{eq:tc_protocol}, where $\hat{H}_1(t)$ denotes the spin-flip operation [see Eq.~\eqref{eq:spinflip:hamilt}] and $\hat{H}_2(t)$ is the operator consisting of the flat-band drive and periodic drive for DMBL, which can be expressed as:
	\begin{equation}
		\hat{H}_2(t) = \hat{H}_{\mathrm{FB}}(t) + \hat{H}_{\mathrm{DMBL}}(t),
	\end{equation}
	where $\hat{H}_{\mathrm{FB}}(t)$ is taken from Eq.~\eqref{eq:fb_t} and $\hat{H}_{\mathrm{DMBL}}(t)$ is taken from Eq.~\eqref{eq:tc_protocol_17}. The time evolution operator for this mixed protocol is illustrated in Fig.~\ref{fig:mixed_protocol}.	

	We numerically simulate the time evolution of the parity adjusted magnetization order parameter $Z(nT) = (-1)^n \langle \hat{\sigma}^z(nT) \rangle$ under the mixed drive combined by flat band and DMBL, for a spin-1/2 chain with sizes $N=9,11,12$ up to 100T, considering nearest-neighbor interactions (TFIM case, where $\beta = \infty$) and all-to all interaction (LMG case, where $\beta =0$). The parameters are set to $\lambda_0 J = J_0 = 0.18$, with the DMBL drive tuned to its freezing condition $\mathcal{J}_0\left(\frac{4h}{\omega}\right)=0$ and frequency $\omega=20$. Additionally, we introduce a spin-rotation error $\varepsilon_r\in\{0,0.01,0.02,0.03\}$ during the $T_1$ spin-flip interval. For the TFIM drive, we consider both open and periodic boundary conditions, whereas for the flat-band drive we restrict to periodic boundary conditions only, since the flat-band protocol is insensitive to boundary effects (see Appendix~\ref{app:boundary_condition}).

	\begin{figure}[t]
		\centering
		\includegraphics[width = 1.0\linewidth]{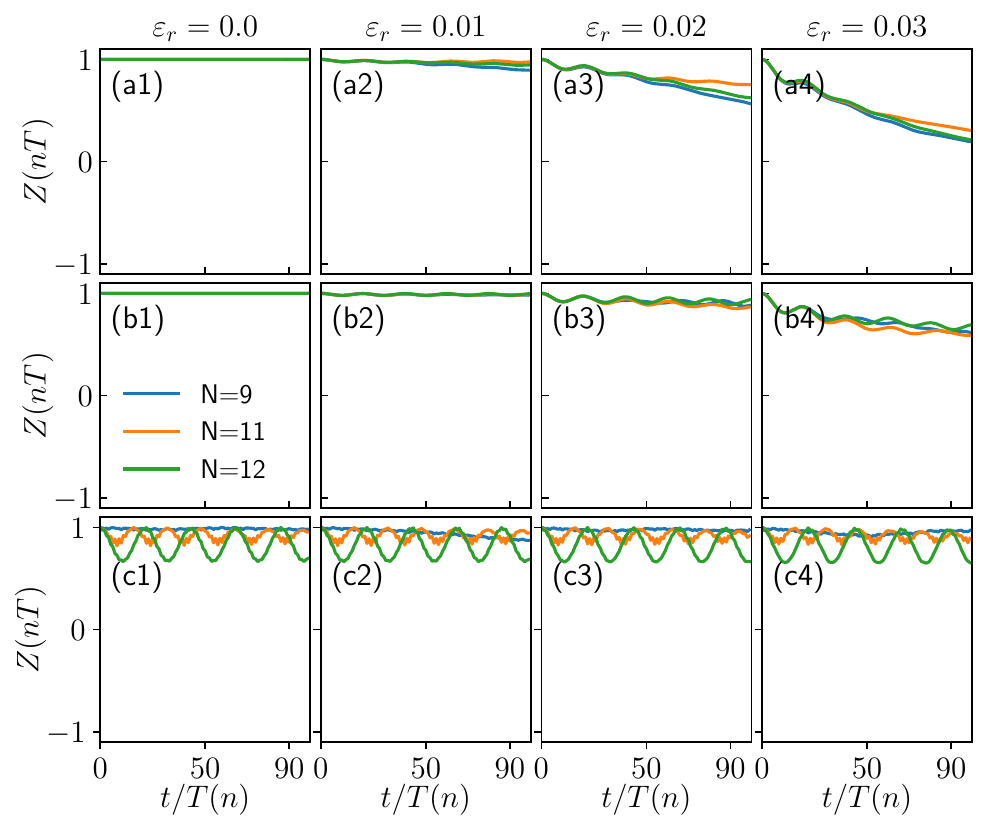}
		\caption{Time evolution of the parity-adjusted magnetization order parameter $Z(nT)=(-1)^n\langle \hat{\sigma}^z(nT)\rangle$ under the mixed protocol for system sizes $N=9,11$ and $12$ up to $100T$. (a1)-(c1), (a2)-(c2), (a3)-(c3), and (a4)-(c4) correspond to rotational errors $\varepsilon_r= 0, 0.01, 0.02$ and $0.03$ respectively in the spin flip operation. For the TFIM-FB mixed case, Periodic boundary conditions (PBC) are applied to the flat-band drive, while both open (OBC) and periodic (PBC) boundary conditions are used for the DMBL drive. (a1)-(a4) show OBC results, and (b1)-(b4) show PBC results. The LMG-FB mixed drive induced DTC case is illustrated in (c1)-(c4).	Drive parameters are set to $\lambda_0 J=J_0=0.18$, $\omega=20$, and $\mathcal{J}_0\left(\frac{4h}{\omega}\right)=0$.}
		\label{fig:mixed_protocol_DTC}
	\end{figure}
	We observe in Fig.~\ref{fig:mixed_protocol_DTC} that the stability of the DTC phase increases with system size and remains robust when TFIM under PBC is combined with the flat-band protocol, in the presence of finite spin-rotation errors ($\varepsilon_r \le 0.02$). In contrast, the DTC phase emerging from TFIM under OBC mixed with the flat-band protocol exhibits a rapid decay of the order parameter at finite $\varepsilon_r$ for all system sizes. This behavior originates from edge effects in the OBC TFIM drive, where edge spins lack neighboring spins on one side. Consequently, OBC removes one stabilizing bond per edge spin, locally reducing the many-body gap that protects the dynamics against spin-rotation errors. The edge then serves as a nucleation site where finite $\varepsilon_r$ first overcomes localization, generating a domain wall that propagates into the bulk. This mechanism is absent under PBC, where translational symmetry provides uniform protection across all sites. These results demonstrate that the boundary conditions in the TFIM--FB mixed drive play a central role in determining DTC robustness in the presence of imperfections.
	
	Additionally, the DTC phase arising from the DMBL in the LMG case, when combined with the flat-band protocol, becomes increasingly unstable with system size for all values of $\varepsilon_r$. This behavior reflects the intrinsic instability of the DMBL phase in the thermodynamic limit~\cite{Rahaman_2024_prb}.
	
	Therefore, the mixed protocol combining the flat-band drive with the DMBL drive under PBC does not exhibit robustness comparable to that of the modified flat-band protocol at higher rotational errors (up to $\varepsilon_r = 0.03$) in presence of additional interactions as described in Sec.~\ref{sec:add_interaction}. This suggests that the mixed protocol can enhance DTC stability against spin-rotation errors, its effectiveness is limited relative to the modified flat-band protocol with additional interactions, it can be utilized to stable the DTC. 
	
	\section{\label{sec:level7}Summary and Outlook}
	\label{sec:conclusion}
	In this work, we have introduced a novel framework for realizing DTC phases in clean spin-1/2 chains by combining a global spin-flip operation with a flat-band drive protocol. The flat-band protocol is meticulously designed to produce a fully degenerate quasienergy spectrum, thereby localizing the system's initial state and suppressing thermalization. This is achieved through the application of two distinct periodic drives, ensuring that the time evolution operator reduces to the identity at each discrete time period. The resulting dynamics stabilizes the periodic flipping of spins, leading to robust period-doubling behavior that breaks intrinsic time-translation symmetry, thereby enabling the emergence of a DTC phase.
	
	Our analysis reveals that the flat-band-induced DTC phase exhibits remarkable stability across a wide range of spin-spin interaction strengths and interaction ranges. However, it demonstrates sensitivity to spin rotational errors, which manifest as rapid oscillations and beat patterns in the dynamics. However, the DTC phase persists for small deviations of the drive parameters from the ideal flat-band condition, as evidenced by high fidelity values. At stroboscopic times, the time evolution operator reduces to the identity, a hallmark of the flat-band protocol. Although this property primarily implies that the flat-band protocol does not induce dynamics, over sufficiently long times, we find that a crossover to a thermal regime can occur due to spin-spin interactions with various interaction ranges. This leads to the emergence of a prethermal DTC phase, enabled by the flat-band protocol, which is distinct from trivial period-doubling phases.
	
	We further conducted a comparative study of the stability of the flat-band-induced DTC phase against those stabilized by disorder-induced MBL and disorder-free DMBL. The results indicate that the flat-band-induced DTC phase is more susceptible to spin rotational errors than the MBL- and DMBL-induced DTC phases. Among these, the DMBL-induced DTC phase demonstrates superior robustness to spin rotational errors relative to the transverse field Ising model (TFIM)-induced DTC phase. Additionally, the flat-band-induced DTC phase exhibits resilience to variations in interaction strengths and ranges, unlike the MBL- and TFIM-induced DTC phases, which are significantly influenced by these parameters~\cite{Khemani_2016_prl, Randall_2021, yao_2017, Pizzi_2021}. Notably, the flat-band-induced DTC phase remains independent of system size, offering a distinct advantage over the MBL- and DMBL-induced DTC phases, particularly in the thermodynamic limit.
	
	To further enhance the robustness of the DTC phase against spin rotational errors, we explored a modified flat band protocol that incorporates an additional spin-spin interaction term which introduces an energy gap in the pure flat-band and subsequently suppresses the accumulation of phase errors from spin rotations. This modification significantly improves the stability of the DTC phase, particularly in regimes characterized by strong interactions and high-frequency driving, as well as in regimes with weak interactions and low-frequency driving. 
    As an alternative stabilization strategy, we investigate a hybrid mixed protocol that interleaves the flat-band and DMBL drives. The resulting DTC exhibits a moderate enhancement of stability relative to the pure flat-band protocol.
    These results demonstrate that the flat-band protocol can be engineered to mitigate the effects of spin-rotation errors, thus improving the robustness of the DTC phase under realistic conditions.
	
	\emph{Outlook.} The proposed flat-band-induced DTC framework paves the way for exploring nonequilibrium phases in clean spin systems. Its reliance on engineered drive sequences, rather than disorder, makes it particularly suitable for implementation on Noisy Intermediate-Scale quantum devices (NISQ)~\cite{Preskill_2018}. Experimental platforms such as trapped ions~\cite{Zhang_2017}, superconducting qubits~\cite{Khemani_2021}, and Rydberg atom arrays~\cite{Browaeys_2020} are well-suited to implement this protocol. For instance, in trapped ions, the global spin-flip operation and flat-band drive can be realized using laser-induced spin-dependent forces and tailored pulse sequences. In superconducting qubits, microwave control fields can generate the required time-dependent Hamiltonians, while Rydberg atom arrays can utilize optical tweezers and Rydberg blockade mechanisms to engineer the desired interactions. However, achieving effective suppression of the system dynamics requires a deeper understanding of the mechanisms governing the stability of the DTC phase, including the potential role of disorder in mitigating the faster oscillating components of the flat-band protocol~\cite{tista2025}.
	
	Although the exact flat-band-induced DTCs phase does not achieve the same level of robustness as the MBL-induced DTC phase in the presence of spin rotational errors~\cite{yao_2017}, it offers a versatile and experimentally accessible route to realization of DTCs in clean driven systems. In contrast to DMBL-induced time crystals, which require fine-tuning of the drive parameters to achieve localization and subharmonic response~\cite{Rahaman_2024_prb}, the flat-band protocol provides a wider parameter space for stabilizing the DTC phases. This flexibility makes it particularly advantageous for experimental platforms where disorder is challenging to implement or control.
	
	In conclusion, the flat-band protocol significantly broadens the scope of time crystal realization, offering a robust and scalable approach to achieving DTC phases in clean, disorder-free systems. Future research could explore the interplay between drive engineering, interaction range, and error resilience, potentially uncovering new regimes of nonequilibrium quantum order.
	
	\begin{acknowledgments}
		MR thanks DST, India, for support through the DST/FFT/NQM/QSM/2024/3 project. AR acknowledges support from the University Grants Commission (UGC) of India, Grant No. F.30-425/2018(BSR), as well as from Anusandhan National Research Foundation (ANRF, formerly SERB), Grant No. CRG/20l8/004002. Both authors express their sincere appreciation to Dr.\ Sayan Choudhury of the Harish-Chandra Research Institute for insightful discussions and valuable feedback on the manuscript.
	\end{acknowledgments}
	
	\section*{Author Contributions}
	\noindent\insertcreditsstatement
	
	\appendix
	\section{Emergence of Flat band}
	\label{app:fb}
	Using Eq.~\eqref{eq:two_drive_protocol_1}, the symmetry inherent in the periodic drive dictates that 
	\begin{equation}
		\label{eq:lambda_symmetry}
		\lambda_{1,2}(\alpha_1 T + t^\prime) = - \lambda_{1,2}(\alpha_1 T - t^\prime),
	\end{equation}
	for $t^\prime \le (\beta_2T - \beta_1 T)/2$. This ensures that for all $t^\prime$, 
	\begin{equation}
		\label{eq:fb_symmetry}
		\hat{H}_{\mathrm{FB}}(\alpha_1 T + t^\prime) = - \hat{H}_{\mathrm{FB}}(\alpha_1 T - t^\prime).
	\end{equation}
	The time evolution operator $\hat{U}(t)$, governed by the Schr\"odinger equation $\displaystyle i \hbar \partial_t \hat{U}(t) = \hat{H}(t) \hat{U}(t)$, can be expressed as
	\begin{align}
		\label{eq:ut0}
		&\hat{U}(T, 0) = \mathcal{T} \exp(-i\int_0^T \hat{H}(t'') dt''),\nonumber\\
		&= \mathcal{T} \exp(-i\int_{T_1}^T \hat{H}_{\mathrm{FB}}(t'') dt'')  \exp(-i\int_{0}^{T_1} \hat{H}_1(t'') dt''),
	\end{align}
	where $\mathcal{T}$ denotes the time-ordering operator. Given the high drive frequencies, we can apply the first-order Suzuki-Trotter decomposition (STD)~\cite{Hatano2005} to approximate the time evolution operator as
	\begin{multline}
		\hat{U}(T, 0) \approx \prod_{m=0}^{N} \left\{\exp(-i \hat{H}_\mathrm{FB}(t_m)\Delta t) + \hat{\mathcal{O}}(\Delta t^2)\right\}\\        
		\times \exp(-i \lambda_s (1-\varepsilon_r)\sum_i \hat{\sigma}^x_i),
	\end{multline}
	where $\Delta t = (T-T_1)/(1+N)$. Given the high effective frequency of the flat-band drive ($2\omega$), we can utilize the Zassenhaus' formula \cite{Fernando_2012, Dupays_2023, Kimura_2017, Rahaman_2024} for two operators $\hat{a}$ and $\hat{b}$, which can be expressed as:
	\begin{equation}
		e^{\hat{a} + \hat{b}} = e^{\hat{a}} e^{\hat{b}} \prod_{n=2}^{\infty} e^{P_n(\hat{a}, \hat{b})},
	\end{equation}
	where $P_n(\hat{a}, \hat{b})$ represents the polynomial $n$-th of $\hat{a}$ and $\hat{b}$. Consequently,
	\begin{equation}
		\label{eq:zassenhaus}
		e^{T_1(\hat{a} + \hat{b})} = e^{T_1\hat{a}} e^{T_1\hat{b}} e^{P_2T_1^2\comm{\hat{a}}{\hat{b}}} e^{P_3T_1^3\comm{\hat{a}}{\comm{\hat{a}}{\hat{b}}}} \dots \approx e^{T_1\hat{a}} e^{T_1\hat{b}},
	\end{equation}
	where higher-order terms $\hat{\mathcal{O}}(T_1^2)$ can be neglected. Applying the Zassenhaus formula in Eq.\eqref{eq:zassenhaus} to the time evolution operator, we obtain
	\begin{align}
		\hat{U}(T, T_1) \equiv \prod_{m=0}^{N} \exp[-i \hat{H}_\mathrm{FB}(t_m)\Delta t].
	\end{align}
	With proper time ordering, the time evolution operator can be written as
	\begin{equation}
		\hat{U}(T, T_1) = \prod_m \hat{U}(t_{m+1}, t_m) = \prod_m \hat{U}_m.
	\end{equation}
	The evolution operator $\hat{U}(T, T_1)$ operates between the turning points $\beta_1$ and $\beta_2$. The null point $\hat{H}(\alpha_1 T) = 0$ implies that $\hat{U}(\alpha_1 T_2 + \Delta t, \alpha_1 T_2) = \mathbf{1}$. By applying the STD along with the Zassenhaus' formula, we can approximate the time evolution operator $\hat{U}(T, T_1)$ as follows:
	\begin{widetext}
		\begin{multline}
			\label{eq:u_FB}
			\hat{U}(T, T_1) \equiv \hat{U}(\beta_2 T_2, \beta_2 T_2 - \Delta t) \hat{U}(\beta_2 T_2 - \Delta t, \beta_2 T_2 - 2\Delta t) \dots \\
			\dots \hat{U}(\alpha_1 T_2 + 2\Delta t, \alpha_1 T_2 + \Delta t) \hat{U}(\alpha_1 T_2, \alpha_1 T_2 - \Delta t) \dots \hat{U}(\beta_1 T_2 + 2\Delta t, \beta_1 T_2 + \Delta t) \hat{U}(\beta_1 T_2 + \Delta t, \beta_1 T_2).
		\end{multline}
	\end{widetext}
	It should be noted that the extreme terms in Eq.~\eqref{eq:u_FB} are exact Hermitian conjugates of each other, as indicated by Eq.~\eqref{eq:fb_symmetry}. This results in:
	\begin{equation}
		\label{eq:flatband_unitary}
		\hat{U}(T, T_1) = \mathbf{1}.
	\end{equation}
	Therefore, the unitary evolution operator becomes:
	\begin{equation}
		\hat{U}(T, 0) = \exp(-i \hat{H}_{\text SF} T_1).
	\end{equation}
	Furthermore, the Hamiltonian $\hat{H}(t)$ is time periodic, and therefore does not conserve energy. Using Floquet theory, one can explain the dynamics of the system in terms of $k$'th quasistationary eigenstates ($\Phi^k(n)$) and the corresponding quasistationary eigenvalues($\Phi^k(E)$), thereby transforming the time dependent problem to time independent problem. The system propagator is derived from the Schr\"odinger equation, $\displaystyle i \hbar \partial_t \hat{U}(t) = \hat{H}(t) \hat{U}(t)$. The time evolution of the system is governed by the Floquet Hamiltonian, which is a time-independent operator constructed from the time-dependent Hamiltonian of the system.
	The floquet operator is defined as
	\begin{equation}
		\hat{\mathcal{F}} = \exp(-i\hat{H}_F T),
	\end{equation}
	here, $\hat{H}_F$ is the Floquet Hamiltonian and $T$ is the drive time period. The Floquet Hamiltonian is derived from the time-dependent Hamiltonian of the system $\hat{H}(t)$, as, $\displaystyle  \hat{H}_F = \left(\hat{H}(t) -i \partialderivative{t}\right)$, evaluated in the stroboscopic time interval $t=nT\; ;n=1,2,3,\dots$.
	
	Taking the Floquet operator as $\hat{\mathcal{F}}$, the effective Floquet Hamiltonian ($\hat{H}^{\mathrm{eff}}$) for the proposed model in $T$ time is given by
	\begin{align}
		\hat{\mathcal{F}} &= \exp(- i \hat{H}^{\mathrm{eff}} T)\nonumber\\
		& = \exp(-i\hat{H}_2 T_2) \exp(-i\hat{H}_1 T_1)
	\end{align}
	
	To understand the system's dynamics during the second time interval, governed solely by the flat-band protocol, we can disregard the spin-flip operation ($\hat{H}_1(t)$) in Eq.~\eqref{eq:tc_protocol}. This simplifies Eq.~\eqref{eq:ut0} to:
	\begin{equation}
		\hat{U}(T, 0)_{\mathrm{FB}} \equiv \mathcal{T} \exp(-i \int_{T_1}^T \hat{H}_{\mathrm{FB}}(t'') dt'') = \mathbf{1}.
	\end{equation}
	Thus, the reduced Floquet operator in the absence of the spin-flip drive is given by:
	\begin{align}
		\label{eq:fb_floq}
		&\hat{\mathcal{F}} = \exp(-i \hat{H}^{\mathrm{eff}} T) = \exp(-i \hat{H}_{\mathrm{FB}} T_2),\\
		\label{eq:fb_floq_2} \mathrm{and}, \;&\hat{\mathcal{F}} = \hat{U}(T, 0)_{\mathrm{FB}} = \mathbf{1}.
	\end{align}
	These equations, Eqs.~\eqref{eq:fb_floq}-\eqref{eq:fb_floq_2}, imply that $\hat{H}^{\mathrm{eff}} = 0$, leading to quasienergy $\Phi^k(E) = 0$ for all $n$'th Floquet states, thereby manifesting a completely degenerate \textit{flat} band. 

	\begin{figure*}[t]
		\centering
		\includegraphics[width=1.0\linewidth]{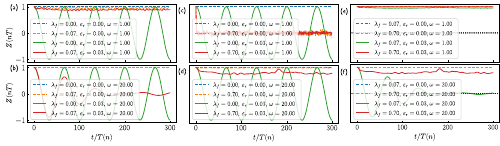}
		\caption{Comparative stability analysis of the DTC phase under the modified flat-band protocol. (a)-(d) present results incorporating flat band and an additional dc interaction term ($\lambda_fJ$), while (e) and (f) show results only due to the interaction term ($\lambda_fJ$). The analysis spans two interaction regimes---strong interactions [$\lambda_0 = 1$, (c) and (d)] and weak interactions [$\lambda_0 = 0.1$, (a) and (b)]---across both high-frequency ($\omega = 20$) and low-frequency ($\omega = 1$) driving regimes. Compared to Fig.~\ref{fig:perfect_flatband}, the modified protocol demonstrates qualitatively distinct DTC behavior in presence of additional interactions [(a)-(d)] and absence of flat band [(e) and (f)], corroborating the emergent stable DTC phase in presence of mutual interaction between flat band and additional interaction term. All simulations were performed for a spin chain of size $N=8$, with parameters fixed at $\lambda_f = 0.7\lambda_0$, $J=1$ and $\omega_0 = 0.0$.}
		\label{fig:compare}
	\end{figure*}
	\section{Flat-band Protocol: In presence of additional Interaction}
	\label{app:interaction_flatband}
	The exact flat-band protocol outlined in Sect.~\ref{sec:model} yields perfectly unitary dynamics, independent of system size and interaction strength. As demonstrated in Sect.~\ref{sec:imperfect_flatband}, this protocol exhibits robustness against small perturbations in the two-tone drive frequencies. However, the introduction of considerable spin-rotation errors during the $T_1$ interval can induce rapid oscillations and beat patterns in the system's dynamics, potentially undermining localization and the subharmonic response. This sensitivity arises from the intrinsic mechanism of the flat-band protocol, which achieves dynamical suppression by engineering a fully degenerate quasienergy spectrum. To harness the flat-band protocol for stabilizing the DTC phase in the presence of such rotational errors, it is necessary to modify the Hamiltonian to maintain an effective spin-spin interaction term that can counteract the effects of spin-rotation imperfections (which is generic in MBL induced DTC). Accordingly, we introduce an additional aperiodic spin-spin interaction term to the two-tone flat-band Hamiltonian from Eq.\eqref{eq:tc_protocol}, as described below.
	\begin{equation}
		\tilde{H}_{\mathrm{FB}}(t) = \lambda_1(t) \sum_i \hat{\sigma}^x_i \hat{\sigma}^x_{i+1} + \lambda_2(t) \sum_i \hat{\sigma}^z_i + \lambda_f J \sum_i \hat{\sigma}^z_i \hat{\sigma}^z_{i+1},	
		\label{eq:fb_interaction_tilde}
	\end{equation}
	where, the time-dependent coefficients are redefined as:
	\begin{align}
		\lambda_1(t) &= +\lambda_0 && ;\frac{mT}{4} < t \le \frac{(m+1)T}{4},\nonumber\\
		&= -\lambda_0 && ;\frac{(m+1)T}{4} < t \le \frac{(m+2)T}{4},\nonumber\\
		\lambda_2(t) &= \omega_0 - \omega_1 && ;\frac{qT}{4p} < t \le \frac{(q+1)T}{4p},\nonumber\\
		&= \omega_0 + \omega_1 && ;\frac{(q+1)T}{4p} < t \le \frac{(q+2)T}{4p},
	\end{align}
	Here, $\lambda_f$ represents a dc-offset to the spin-spin interaction strength, and for analytical convenience, the time coefficients are fixed at $m, q =0,2,4, \dots$. Here $`p$' can be any integer, for simplicity, we set it to $p=2$. Additionally, we consider $\omega_0 =0$. The time evolution operator $\hat{U}(T, 0)$ for this modified Hamiltonian can be expressed as:
	\begin{equation}
		\hat{U}(T, 0) = \mathcal{T} \left(\hat{U}_4 \hat{U}_3 \hat{U}_2 \hat{U}_1 \hat{U}_0\right),
	\end{equation}
	where the unitary operators $\hat{U}_i$ for $i = 0, 1, 2, 3$ and $4$ are defined as
	\begin{align}
		&\hat{U}_0 = \exp(-i \hat{H}_0\frac{T}{2}),\quad \hat{H}_0 = \lambda_s (1-\varepsilon_r) \sum_i \hat{\sigma}^x_i,\\
		&\hat{U}_1 = \exp(-i \hat{H}_1\frac{T}{8}), \quad \hat{H}_1 = \lambda_f  \hat{A}-\lambda_0 \hat{B} + \omega_1 \hat{C},\\
		&\hat{U}_2 = \exp(-i \hat{H}_2\frac{T}{8}),\quad \hat{H}_2 = \lambda_f  \hat{A}-\lambda_0 \hat{B} -\omega_1 \hat{C},\\
		&\hat{U}_3 = \exp(-i \hat{H}_3\frac{T}{8}),\quad \hat{H}_3 = \lambda_f  \hat{A}+\lambda_0 \hat{B} + \omega_1 \hat{C},\\
		&\hat{U}_4 = \exp(-i \hat{H}_4\frac{T}{8}),\quad \hat{H}_4 = \lambda_f  \hat{A} +\lambda_0 \hat{B} -\omega_1 \hat{C},
	\end{align}
	and $\hat{A} = \sum_i J\hat{\sigma}^z_i \hat{\sigma}^z_{i+1}, \hat{B} = J\sum_i \hat{\sigma}^x_i \hat{\sigma}^x_{i+1}$, and $\hat{C} = \sum_i \hat{\sigma}^z_i$.
	We apply the Magnus expansion~\cite{Kuwahara_2016} to approximate the time evolution operator as
	\begin{align*}
		\hat{U}(T, 0) =& \mathcal{T} \exp\left(\int_0^T dt  \hat{H}(t)\right)\\ =& \exp(-i \hat{H}_{\mathrm{ME}} T),
		\label{eq:unitaryme}
	\end{align*}
	where $\hat{H}_{\mathrm{ME}}$ is the effective Hamiltonian derived from the Magnus expansion, which provides this as a series;
	\begin{equation}
		\hat{H}_{\mathrm{ME}} = \sum_{n=0}^{\infty} T^n \xi_n.
	\end{equation}
	Taking into account the fastest drive frequency $\Omega = 2\omega$, we divide the total time period $T$ into five intervals, consisting of one T/2 for the spin flip operation and four of equal duration $T/8$. The Hamiltonians in each interval are denoted as $H_i$ for $i = 0, 1, 2, 3$ and $4$. The Magnus expansion terms can be computed using these Hamiltonians.
	
	\noindent The first-order term of the Magnus expansion is given by:
	\begin{align}
		\xi_0 =& \frac{1}{T} \int_0^T \hat{H}(t_1) dt_1 \equiv \sum_{k=0}^{2p-1} \hat{H}_k\nonumber\\
		=&\frac{1}{T} \Big\{\hat{H}_0 \frac{T}{2} + (\hat{H}_1 + \hat{H}_2 + \hat{H}_3 + \hat{H}_4)\frac{T}{8} \Big\}.
	\end{align}
	Calculating this, we find the following.
	\begin{equation}
		\xi_0 = \frac12 \left(\lambda_s (1-\varepsilon_r) \sum_i \hat{\sigma}^x_i + \lambda_f J \sum_i \hat{\sigma}^z_i \hat{\sigma}^z_{i+1} \right).
	\end{equation}
	The second-order term is given by:
	\begin{equation}
		\xi_1 = \frac{i}{2T} \sum_{i>j} \comm{\hat{H}_i}{\hat{H}_j}.
	\end{equation}
	Evaluating this term, we obtain the following.
	\begin{equation}
		\xi_1 = -\frac{\lambda_s (1-\varepsilon_r)\lambda_f J T^2}{4}\sum_i \left(\hat{\sigma}^y_i\hat{\sigma}^z_{i+1} +\hat{\sigma}^z_i\hat{\sigma}^y_{i+1} \right).
	\end{equation}
	For small offsets and transverse field oscillation amplitudes, we may assume that $\lambda_f J< \omega_1 < \lambda_0$. Thus,
	\begin{widetext}
		\begin{equation}
			\hat{H}_{\mathrm{ME}} = \frac{\lambda_f J}{2} \sum_i \hat{\sigma}^z_i \hat{\sigma}^z_{i+1}+ \frac{\lambda_s(1-\varepsilon_r)}{2} \sum_i \hat{\sigma}^x_i -\frac{\lambda_s (1-\varepsilon_r)\lambda_f J T}{4}\sum_i \left(\hat{\sigma}^y_i\hat{\sigma}^z_{i+1} +\hat{\sigma}^z_i\hat{\sigma}^y_{i+1} \right) + \dots.
			\label{eq:ki_0}
		\end{equation}
	\end{widetext}
	We can express the effective Hamiltonian in generic form as follows.
	\begin{equation}
		\hat{H}_{\mathrm{ME}} \equiv \frac{\lambda_f J}{2} \sum_i \hat{\sigma}^z_i \hat{\sigma}^z_{i+1}+ \frac{\lambda_s(1-\varepsilon_r)}{2} \sum_i \hat{\sigma}^x_i + \mathcal{\hat{O}}(T).
		\label{eq:ki}
	\end{equation}
	Here, $\displaystyle \mathcal{\hat{O}}(T)$ denote the leading-order correction arising from the noncommuting nature of the drive components. In Eq.~\eqref{eq:ki}, the first term encodes the additional spin-spin interaction, while the second term realizes the spin-flip operation; since $\lambda_s T = \pi$, the latter is frequency independent. Unitary evolution is governed by $\displaystyle \hat{U}(T, 0) = \exp(-i \hat{H}_{\mathrm{ME}} T)$, where the effective spin interaction strength is characterized by $\eta \equiv \frac{\lambda_fJ}{\omega}$. The mitigation effect of spin-rotation errors requires $\eta$ to be large enough to compensate for the melting of the DTC order for finite $\varepsilon_r$. Furthermore, for typical parameter regimes where $\lambda_fJ, \omega_0 \ll \lambda_0, \omega_1$, the correction term $\mathcal{\hat{O}}(T)$ remains negligible.
	
	In the high-frequency regime, the correction term is strongly suppressed, and the stability of the DTC under finite $\varepsilon_r$ is governed by the competition between the interaction and spin-flip terms. When interactions are weak, $\eta$ is negligible, and the imperfect spin-flip term dominates, inducing beating in the temporal order parameter and ultimately destabilizing the DTC phase. In contrast, strong interactions ($\eta \gg 1$) effectively counteract rotational imperfections and preserve the phase.
	
	Reducing the frequency causes the correction term $\mathcal{\hat{O}}(T)$ to become appreciable. For strong interactions, its nontrivial interplay with the spin-flip and interaction terms generates complex dynamics that undermine DTC stability. In the weak-interaction regime, however, the correction remains negligible due to small interaction terms. Importantly, at low frequencies, even weak interactions produce $\eta \gg 1$ at the small denominator $\omega$, allowing the interaction term to compensate for spin-rotation errors and maintain a robust subharmonic response.
	
	Nevertheless, a more comprehensive investigation is needed to determine how the additional interaction term and the correction term jointly govern DTC stability across the full parameter space, despite the perfectly unitary dynamics of the exact flat-band protocol. Our results show that the added terms $\lambda_f$ stabilize the DTC phase only in regimes where the correction term in Eq.~\ref{eq:ki} is negligible; once this term becomes appreciable, the robustness of the phase deteriorates. Hence, DTC stability is controlled not by the absolute magnitude of the further added interactions alone, but by their nontrivial interplay with the correction term (see Fig.~\ref{fig:compare}), which can either strengthen or weaken localization and the associated subharmonic response depending on the parameter regime.
	
	To gain deeper physical insight into the stabilization mechanism, we employ \emph{toggling frame analysis}, which decomposes the imperfect global spin flip into its ideal component and the error perturbation. Under this analysis, the system dynamics can be conveniently described by transforming to a toggling frame that rotates by $\displaystyle R_\pi \equiv \Pi_i \exp(-i \pi \hat{\sigma}^x_i)$ after each spin-flip operation~\cite{Choi_2017_prl}. In this toggling frame, the Floquet operator for the modified flat-band protocol reads as
	\begin{equation*}
		\hat{\mathcal{F}}_{\text{Toggle}} = \hat{U}_{\mathrm{FB}^\prime} \hat{U}_{\pi_{{x}}} \exp(-i \lambda_s \varepsilon_r \sum_i \hat{\sigma}^x_i),
	\end{equation*}
	where $\hat{U}_{\mathrm{FB}^\prime} = \exp(-i \hat{H}_{\mathrm{FB}^\prime} T)$ with $\hat{H}_{\mathrm{FB}^\prime} = \hat{H}_{\mathrm{FB}} + \hat{H}_{\mathrm{int}}$, the additional interaction being $\hat{H}_{\mathrm{int}} = \frac{\lambda_f J}{2} \sum_i \hat{\sigma}^z_i \hat{\sigma}^z_{i+1}$, and $\hat{U}_{\pi_{{x}}} = \exp(-i \lambda_s \pi \sum_i \hat{\sigma}^x_i)$ represents the perfect spin-flip operation. The final exponential factor encodes the spin-rotation error.
	
	A crucial insight emerges from examining the parity properties under the $\pi_x$ pulse. Since conjugation by $\hat{\sigma}^x$ transforms operators as $\hat{\mathcal{O}}\mapsto\hat{\sigma}^x\,\hat{\mathcal{O}}\,\hat{\sigma}^x$, we have the transformation rules:
	\begin{equation*}
		\hat{\sigma}^x \xrightarrow{\pi_x} +\hat{\sigma}^x, \quad
		\hat{\sigma}^y \xrightarrow{\pi_x} -\hat{\sigma}^y, \quad
		\hat{\sigma}^z \xrightarrow{\pi_x} -\hat{\sigma}^z.
	\end{equation*}
	The two-tone drive of the ideal flat-band protocol is odd under this transformation ($\hat{\sigma}^x\,\hat{H}_{\mathrm{FB}}\,\hat{\sigma}^x = -\hat{H}_{\mathrm{FB}}$), whereas the additional interaction term is even ($\hat{\sigma}^x\,\hat{H}_{\mathrm{int}}\,\hat{\sigma}^x = +\hat{H}_{\mathrm{int}}$). This distinct parity structure has profound consequences: in the toggling frame, the ideal flat-band drive generates a zero-energy flat band in the quasienergy spectrum, while the additional interaction term generates structure orthogonal to this flat-band sector. Consequently, $\hat{H}_{\mathrm{int}}$ survives the toggling averaging intact and occupies an orthogonal symmetry sector from the flat-band drive, ensuring that the two contributions commute in their symmetry sectors. The interaction term introduces an energy scale $\sim \lambda_f J$, which effectively counteracts the destabilizing dynamics induced by spin-rotation errors, thereby extending DTC stability to a broader parameter regime.

	Taken altogether, these findings indicate that the flat-band protocol, combined with an additional spin-spin interaction, provides a viable route to stabilizing the DTC phase against spin-rotation errors in two distinct regimes: high-frequency driving with strong interactions and low-frequency driving with weak interactions.

	\section{Boundary condition effects on DTC stability under the flat-band protocol}
	\label{app:boundary_condition}

	\begin{figure}[]
		\centering
		\includegraphics[width = 1.0\linewidth]{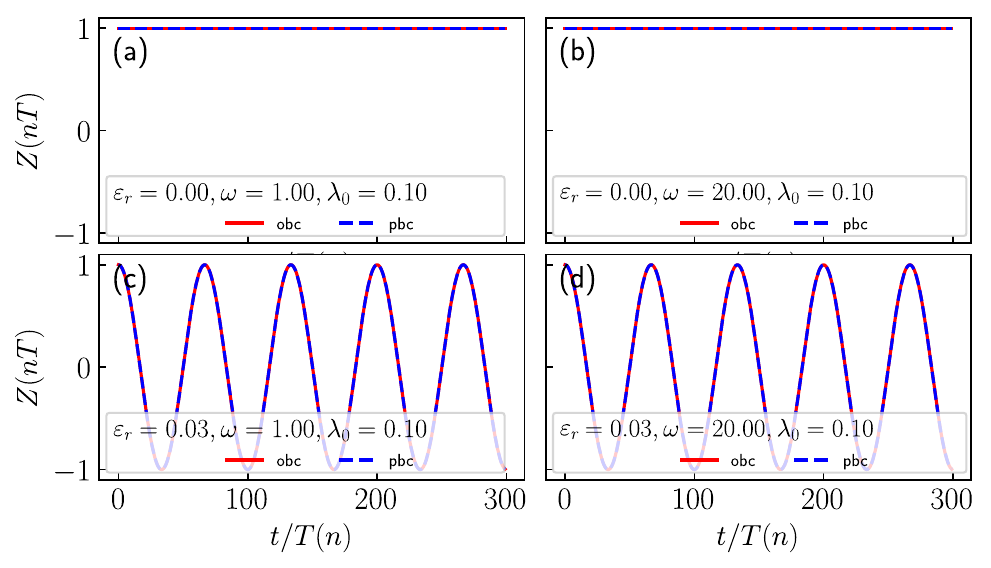}
		\caption{The temporal evolution of the parity adjusted magnetization order parameter $Z(nT) = (-1)^n \langle \hat{\sigma}^z(nT) \rangle$ under the flat-band protocol for both open boundary conditions (OBC) and periodic boundary conditions (PBC). (a) and (b) show results for the ideal flat-band protocol without imperfections ($\varepsilon_r = 0$), while (c) and (d) incorporate spin-rotation errors ($\varepsilon_r = 0.03$). The simulations are performed for a spin-1/2 chain of size $N=8$, with parameters $\lambda_0 J \in [0.1, 1]$ and $\omega \in [1, 20]$.}
		\label{fig:1obcpbc1}
	\end{figure}		
	The unitarity of the stroboscopic time-evolution operator $\hat{U}(T,0)_{\mathrm{FB}}$ in Eq.~\eqref{eq:flatband_unitary} indicates that the flat band arises from an exact cancellation of the evolution at stroboscopic times, a consequence of the drive-protocol symmetries and the noncommutativity of the Hamiltonian terms (can be concluded from Eqs.~\eqref{eq:lambda_symmetry}, \eqref{eq:fb_symmetry}). This result remains valid irrespective of interaction strength, interaction range. However, the choice of boundary conditions for spin interactions can influence system dynamics, particularly in finite-size settings. It is therefore important to investigate how open and periodic boundary conditions affect the stability and localization properties of the DTC phase under the flat-band protocol, especially in the presence of imperfections or additional interactions. Such an analysis provides insight into the robustness of the DTC phase and the role of boundary conditions in shaping the dynamics. To this end, we perform numerical simulations of the time evolution of the parity adjusted magnetization order parameter, $Z(nT) \equiv (-1)^n \langle \hat{\sigma}^z(nT) \rangle$, under the flat-band protocol for both open boundary conditions (OBC) and periodic boundary conditions (PBC).
	\begin{figure}[]
		\centering
		\includegraphics[width = 1.0\linewidth]{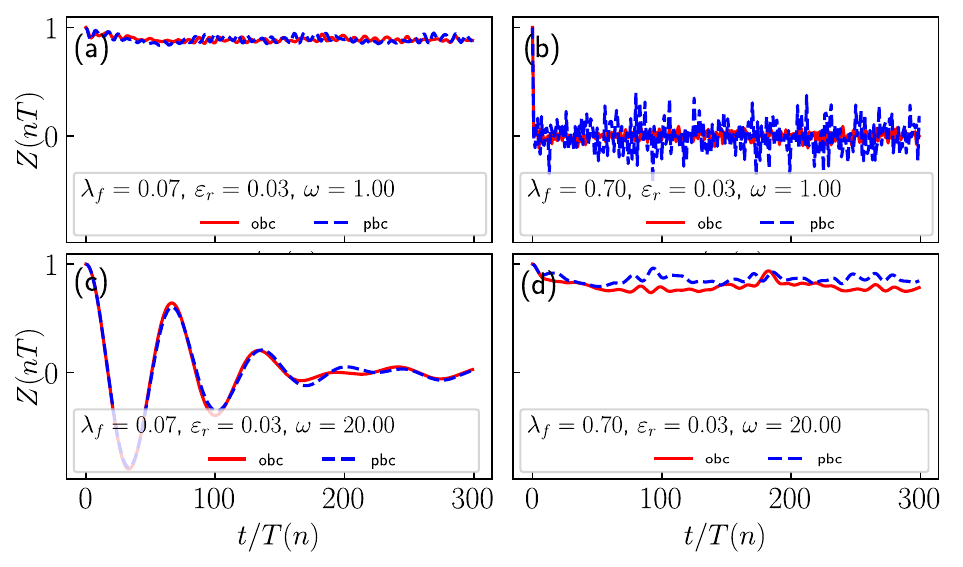}
		\caption{The temporal evolution of the parity adjusted magnetization order parameter $Z(nT) = (-1)^n \langle \hat{\sigma}^z(nT) \rangle$ under the flat-band protocol in presence of spin-rotation errors($\varepsilon_r = 0.03$) for both open boundary conditions (OBC) and periodic boundary conditions (PBC) for the modified flat-band protocol with an additional interaction term. (a) and (d) show results for the modified protocol with $\lambda_f J = 0.7 \lambda_0$ across different frequency regimes ($\omega \in [1, 20]$) and interaction strengths ($\lambda_0 \in [0.1, 1]$). The simulations are performed for a spin-1/2 chain of size $N=8$, with parameters fixed at $\varepsilon_r = 0.03$, $J=1$ and $\omega_0 = 0.0$.}
		\label{fig:1obcpbc2}
	\end{figure}

	We consider a spin-1/2 chain of size $N=8$, with parameters in the ranges $\lambda_0 J \in [0.1, 1]$ and $\omega \in [1, 20]$. To investigate the system dynamics, we numerically compute the time evolution of the parity adjusted magnetization order parameter $\displaystyle Z(nT)$ for both OBC and PBC. Under the ideal flat-band protocol, we first consider the case without imperfections ($\varepsilon_r = 0$), shown in Figs.~\ref{fig:1obcpbc1}(a) and ~\ref{fig:1obcpbc1}(b), and then introduce spin-rotation errors ($\varepsilon_r = 0.03$), shown in Figs.~\ref{fig:1obcpbc1}(c) and ~\ref{fig:1obcpbc1}(d), to assess DTC dynamics under both boundary conditions. We observe that, for the ideal flat-band protocol, the presence or absence of rotational errors does not alter the qualitative behavior of the DTC phase; OBC and PBC yield similar dynamics.
	
	To stabilize the DTC phase against spin-rotation errors, the flat-band protocol requires additional spin interactions [see Eq.~\eqref{eq:fb_interaction}], given by
	\begin{equation*}
		\tilde{H}_{\mathrm{FB}}(t) = \lambda_1(t) \sum_i \hat{\sigma}^x_i \hat{\sigma}^x_{i+1} + \lambda_2(t) \sum_i \hat{\sigma}^z_i + \lambda_f J\sum_i \hat{\sigma}^z_i \hat{\sigma}^z_{i+1}.
	\end{equation*}
	To assess whether the effectiveness of these interactions depends on boundary conditions, we investigate their interplay by considering the spin-1/2 chain with the additional interaction term of strength $\lambda_f J = 0.7 \lambda_0$ across multiple frequency regimes ($\omega \in [1, 20]$) and interaction strengths ($\lambda_0 \in [0.1, 1]$). Numerical simulations exhibits that the DTC phase dynamics under the modified flat-band protocol are qualitatively similar for both OBC and PBC across all explored parameter regimes, as illustrated in Fig.~\ref{fig:1obcpbc2}. This similarity is particularly pronounced in the stable DTC regime ($\eta \gg 1$), as illustrated in Fig.~\ref{fig:1obcpbc2}(a) and ~\ref{fig:1obcpbc2}(d) and confirmed by the discussion in Appendix~\ref{app:interaction_flatband}. While minor quantitative differences are observed in the transient dynamics and the decay rate of the order parameter, the stability of the DTC phase remains robust under both OBC and PBC spin interactions in the parameter space where the DTC phase is stable. These findings indicate that boundary conditions do not significantly affect the stability of the DTC phase under the flat-band protocol, even in the presence of imperfections and additional interactions, thereby demonstrating that the underlying mechanism for DTC stabilization is fundamentally robust against boundary condition variations.

	\bibliography{main}
\end{document}